\pdfoutput=1
\documentclass[%
 reprint,
superscriptaddress,
 amsmath,amssymb,
 aps,
floatfix,
]{revtex4-2}

\usepackage{graphicx}
\usepackage{dcolumn}
\usepackage{bm}
\usepackage{amsthm}

\usepackage[colorlinks]{hyperref}
\usepackage{amsmath,amsfonts,amssymb,color,graphicx,xcolor,physics}
\usepackage{tikz}
\usepackage{float}
\makeatletter
\let\newfloat\newfloat@ltx
\makeatother
\usepackage{algorithm} 
\usepackage{algorithmicx}
\usepackage[noend]{algpseudocode}

\newtheorem{definition}{Definition}
\newcommand\round[1]{\left[#1\right]}

\def\algbackskip{\hskip-\ALG@thistlm}

\begin{document}

\preprint{APS/123-QED}

\title{Tailored XZZX codes for biased noise}

\author{Qian Xu}
\affiliation{Pritzker School of Molecular Engineering, The University of Chicago, Chicago 60637, USA}

\author{Nam Mannucci}
\affiliation{Pritzker School of Molecular Engineering, The University of Chicago, Chicago 60637, USA}

\author{Alireza Seif}
\affiliation{Pritzker School of Molecular Engineering, The University of Chicago, Chicago 60637, USA}

\author{Aleksander Kubica}
\affiliation{AWS Center for Quantum Computing, Pasadena, CA 91125, USA}
\affiliation{California Institute of Technology, Pasadena, California, 91125, USA}

\author{Steven T. Flammia }
\affiliation{AWS Center for Quantum Computing, Pasadena, CA 91125, USA}
\affiliation{California Institute of Technology, Pasadena, California, 91125, USA}

\author{Liang Jiang}
\email{liang.jiang@uchicago.edu}
\affiliation{Pritzker School of Molecular Engineering, The University of Chicago, Chicago 60637, USA}
\affiliation{AWS Center for Quantum Computing, Pasadena, CA 91125, USA}

\date{\today}

\begin{abstract}
Quantum error correction (QEC) for generic errors is challenging due to the demanding threshold and resource requirements. 
Interestingly, when physical noise is biased, we can tailor our QEC schemes to the noise to improve performance. 
Here we study a family of codes having XZZX-type stabilizer generators, including a set of cyclic codes generalized from the five-qubit code and a set of topological codes that we call generalized toric codes (GTCs).
We show that these XZZX codes are highly qubit efficient if tailored to biased noise. 
To characterize the code performance, we use the notion of effective distance, which generalizes code distance to the case of biased noise and constitutes a proxy for the logical failure rate. 
We find that the XZZX codes can achieve a favorable resource scaling by this metric under biased noise. 
We also show that the XZZX codes have remarkably high thresholds that reach what is achievable by random codes, and furthermore they can be efficiently decoded using matching decoders. 
Finally, by adding only one flag qubit, the XZZX codes can realize fault-tolerant QEC while preserving their large effective distance. 
In combination, our results show that tailored XZZX codes give a resource-efficient scheme for fault-tolerant QEC against biased noise. 
\end{abstract}

\maketitle


\section{\label{sec:intro} Introduction}
Quantum error correction (QEC) lies at the heart of robust quantum information processing \cite{nielsen_chuang_2010, lidar_brun_2013}. 
Actively correcting generic errors, such as depolarizing noise, is challenging because error-correcting codes designed for such errors have a  relatively low threshold and require large resource overhead~\cite{fowler2012surface, litinski2019game, chao2020optimization, beverland2021cost}. 
However, physically relevant errors typically have certain structures, which can be exploited to design QEC schemes that are less demanding.
As an example, many physical systems, such as bosonic systems encoded in a so-called cat code~\cite{cochrane1999macroscopically, lescanne2020, grimm2020, mirrahimi2014, puri2020bias}, have a noise channel biased towards dephasing. 
One can then take advantage of the noise bias and design QEC codes that have a boosted performance against the biased noise~\cite{robertson2017tailored, tuckett2018ultrahigh, tuckett2019tailoring, tuckett2020fault, ataides2021xzzx,darmawan2021practical}. 
In particular, Ref.~\cite{ataides2021xzzx} shows that 
the so-called XZZX surface codes with XZZX-type stabilizers exhibit exceptionally high thresholds as well as reduced resource overhead when the noise is biased towards dephasing.

As the error rates of physical systems readily approach or even fall below the fault tolerance threshold \cite{arute2019quantum, egan2020fault, wu2021strong, zhao2021realizing}, it is the resource overhead that ultimately limits the practical application of QEC schemes. 
The analysis of the XZZX surface codes in Ref.~\cite{ataides2021xzzx} showed very promising thresholds for that class of codes, but the question of resource overhead was only briefly addressed.

In this work, we focus on designing QEC codes and schemes that can reduce the resource overhead for fault-tolerant QEC under experimentally relevant biased noise. 
More specifically, given a (finite) noise bias, we aim to find codes that use as few qubits as possible to suppress the logical error rate to a target level. 
As we will show in the Results, instead of numerically extracting the logical error rates, we can 
characterize the performance of different codes against biased Pauli noise by estimating their effective code distance $d^{\prime}$, which takes the bias into consideration and serves as an analogy to the code distance $d$ in the biased-noise setting. 
The notion of effective distance was introduced in Ref.~\cite{dua2022clifford}, and here we use an alternative (though related) definition. 
For the physical error rate $p\ll 1$, the effective code distance $d^\prime$ approximately determines how the logical error rate $p_L$ scales with $p$, i.e., $p_L \sim p^{d^\prime/2}$, and it thus serves as a good proxy for the logical error rate. 
Now our task simply becomes finding codes that use the minimal number of qubits $n$ to reach a target effective distance $d^{\prime}$ among certain code families. 
Given a target effective distance, we can then characterize the efficiency of a code by the code size required to achieve that effective distance.

To construct highly qubit efficient codes, we start from the observation that the well-known five-qubit code \cite{gottesman1997stabilizer}, with stabilizers comprising the cyclic permutation of $XZZXI$, is the smallest code among all possible codes with its effective distance (3 and 5, respectively) for both depolarizing and infinitely biased Pauli Z noise. This indicates that codes with XZZX-type stabilizers could potentially be resource efficient over a wide range of bias~\cite{robertson2017tailored}. We generalize the five-qubit code by introducing a family of XZZX cyclic codes, which inherit the cyclic structure and all have weight-four XZZX-type stabilizers.
These cyclic codes can reach the optimal effective-distance $n = d^{\prime}$ against infinitely biased noise since they exhibit a repetition-code structure under pure Pauli $Z$ noise. 

To facilitate the analysis of their performance under finite-bias noise, we map them to a family of topological codes.
Concretely, by wrapping the cyclic codes around a torus, we find that these codes belong to a family of XZZX generalized toric codes (GTCs), first introduced in Ref.~\cite{kovalev2012improved} (albeit called checkerboard and non-bipartite rotated toric codes). The GTCs are constructed by first drawing a square qubit lattice with faces representing the XZZX stabilizers, and then identifying qubits that differ by a periodicity vector within the span of two basis periodicity vectors $\vec{L}_1, \vec{L}_2$ (see Fig.~\ref{fig:cyclic_codes_to_GTCs}c). A GTC is therefore specified by its periodic boundary condition induced by $\vec{L}_1$ and $\vec{L}_2$. 
The GTCs share similarly high thresholds with the XZZX surface codes, which we attribute to the local equivalence of their check operators on a torus. Using our tailored efficient decoders, the code-capacity thresholds of the GTCs roughly track the Hashing bound (what is achievable with random coding~\cite{bennett1996mixed, wilde2019quantum}), and their phenomenological thresholds increase from $3.5 \%$ to $10\%$ when the bias parameter (which we will introduce later) increases moderately from 1 to 4. 

More importantly, because of the nontrivial boundary conditions (meaning nontrivial choices of $\vec{L}_1, \vec{L}_2$), the GTCs can be more resource efficient than the XZZX surface codes with either the open or closed rectangular boundaries considered in Ref.~\cite{ataides2021xzzx}. 
We derive the effective distance of the GTCs using topological (or geometrical) tools, and from this we can optimally choose $\vec{L}_1, \vec{L}_2$ if given the value of a bias parameter $\omega$ (defined later). 
The optimal codes satisfy $n = d^{\prime 2}/2\omega$, which indicates that the tailored GTCs require resource that scale quadratically in the target effective distance, similarly to the standard surface codes (for depolarizing noise), but with a reduction by a factor of $2\omega$.

Combining the analysis for the cyclic codes and the GTCs, we obtain the optimal performance for the XZZX codes, given a bias parameter $\omega$.
For $n \leq 2\omega$, the optimal codes are those with cyclic (repetition-code) structures and the optimal resource-distance dependence $n = d^{\prime}$ is achieved.
For $n > 2\omega$, the optimal codes are the GTCs with an optimized layout and have a quadratic resource scaling $n = {d^{\prime}}^2/2\omega$ with an extra reduction by the factor of $2\omega$.

Lastly, we show that we can preserve the large effective distance of the tailored XZZX codes and maintain the scaling of the logical error rate $p_L \sim p^{d^{\prime}/2}$ in the fault-tolerant regime by using only \emph{one} flag qubit, which is recently introduced for low-overhead fault-tolerant QEC \cite{chao2018quantum, chao2018fault, chamberland2018flag, reichardt2020fault}.

\section{Results}
\noindent
\textbf{Effective code distance for asymmetric Pauli noise} \\
In this work, we consider error correction under an i.i.d.\ asymmetric Pauli channel $\mathcal{E}(\rho) = (1 - p)\rho + \sum_{\sigma \in \{X,Y,Z\}}p_{\sigma} \sigma \rho \sigma$, where $\{p_{\sigma}\}$ denotes an asymmetric probability distribution of three Pauli errors and $p = \sum_{\sigma}p_{\sigma}$ is the total error probability. 
The largest Pauli error probability is denoted as $p_m$, i.e.\ $p_m = \max_{\sigma}p_{\sigma}$. 
To estimate the performance of different error-correcting codes under the asymmetric channel, we define the effective distance $d^{\prime}$ of a stabilizer code as the minimum modified weight of logical operators, with the noise-modified weight of a Pauli $\sigma$ given by $\textrm{wt}^{\prime}(\sigma) \equiv \log p_{\sigma}/\mathcal{N}$, where $\mathcal{N}$ is a normalization factor;
see Ref.~\cite{dua2022clifford} for an alternative but related definition of the effective code distance.
To normalize the effective weight of the most probable Pauli error to 1, we choose $\mathcal{N} = \log p_m$. 
The effective weight of a $n$-qubit Pauli string $P = \bigotimes_{i = 1}^{N} \sigma_i$ with non-identity support on $N$ qubits characterizes its error probability since, by definition, $\textrm{Pr}(P) = p_m^{\sum_{i = 1}^{N}\textrm{wt}^{\prime}(\sigma_i)}\times (1 - p)^{n - N} \approx p_m^{\textrm{wt}^{\prime}(P)}$ (to leading order in $p$).
The effective code distance, therefore, roughly characterizes how the logical error rate $p_L$ scales with the physical error rate $p$: 
In general $p_L$ is suppressed to certain order $r$ of $p$, i.e.\ $p_L \sim p^{r}$, and $r$ is approximately given by $d^{\prime}/2$. 
Under depolarizing noise, the effective weight of a Pauli operator reduces to the Hamming weight and the effective code distance reduces to the code distance $d$. 
Under infinitely biased noise (pure $\sigma$ noise), the effective weight simply counts $\sigma$ as 1 and other Pauli operators as $\infty$, and the effective distance $d_{\sigma}$ of a code is the minimum Hamming weight of the logical operators consisting of only $\sigma$ and identity. 
Without loss of generality, in the rest of the paper we will consider noise biased towards Pauli $Z$ errors, i.e.\ $p_Z \geq p_X, p_Y$, unless specially noted.

\begin{figure}[ht!]
    \centering
    \includegraphics[width = 0.48\textwidth]{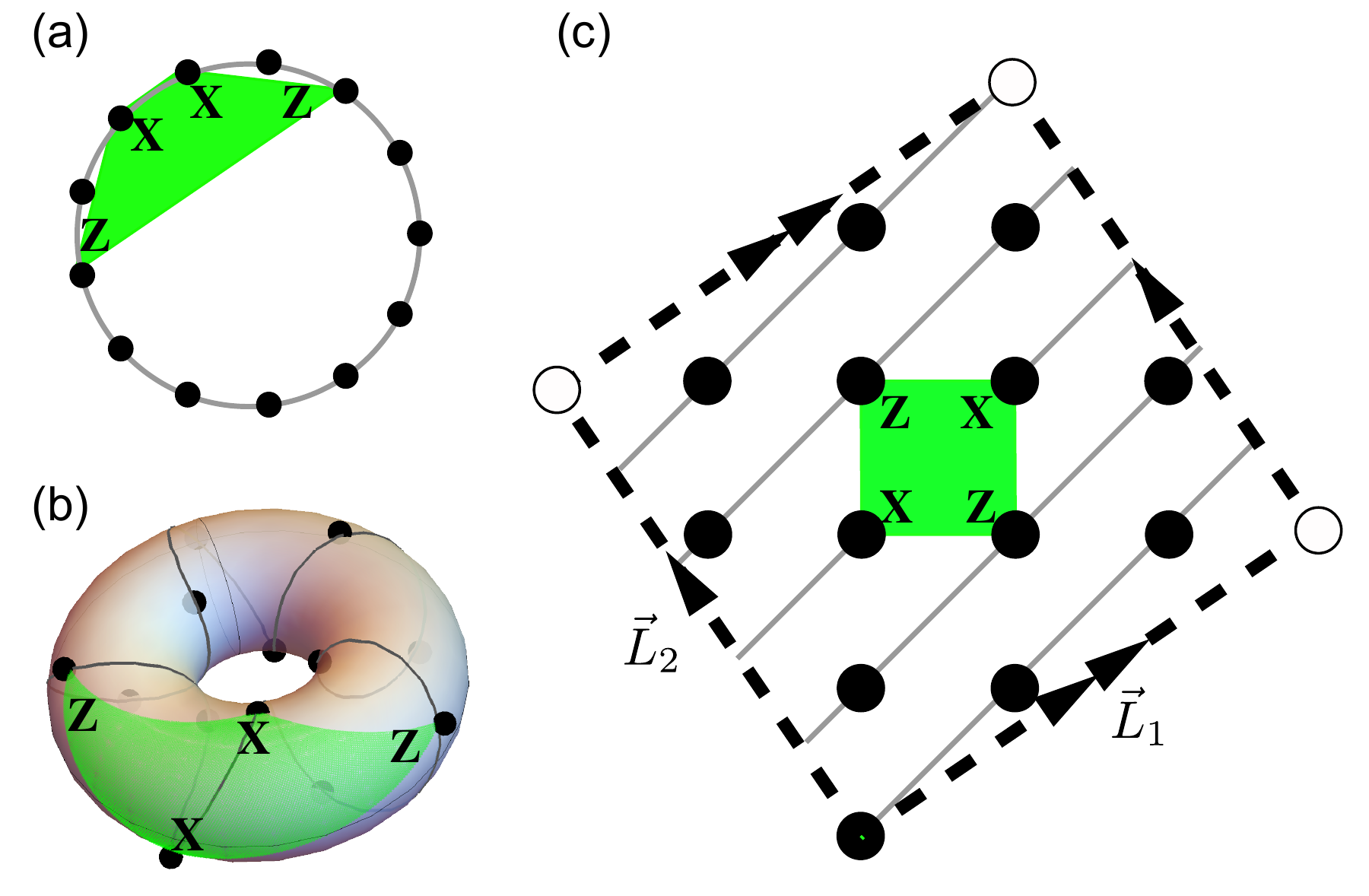}
     \caption{Map from the $\mathcal{S}(13, 2, 1)$ code (a) to a GTC with $\vec{L}_1 = (3,2), \vec{L}_2 = (-2,3)$ (b)(c). 
     The qubits are represented by black dots and labeled along the solid grey string. 
     The stabilizer generators are represented by green plaquettes (only one generator is plotted and others are obtained by shifting along the grey string). 
     (b) and (c) are obtained by wrapping the grey string around the torus. 
     (c) is the 2D layout of (b), with opposite sides of the parallelogram enclosed by $\vec{L}_1$, $\vec{L}_2$ identified.}
    \label{fig:cyclic_codes_to_GTCs}
\end{figure}

The introduction of the effective code distance greatly clarifies our task. 
We simply aim to find codes that can reach a large effective code distance using a small number of physical qubits. 
Given a family of codes, we optimize them by finding the codes that reach a given effective distance $d^{\prime}$ with a minimum code size $n$, or equivalently the codes that can reach the largest effective distance $d^{\prime}$ given a code size $n$.

~\\
\noindent
\textbf{The XZZX cyclic codes} \\
The five-qubit code, with stabilizers generated by cyclic permutations of $ZXXZI$, is the smallest code with distance $d = 3$.
Moreover, it is also the smallest code with $d_{Z} = 5$, where $d_{Z}$ denotes the effective distance under pure Pauli $Z$ noise, since it exhibits a repetition-code structure for pure Pauli $Z$ noise.
Therefore, it is a resource efficient code for both depolarizing and infinitely biased Pauli $Z$ noise.
To find larger codes that are efficient under a wide range of noise bias, we can generalize the five-qubit code and consider a family of XZZX cyclic codes with stabilizer groups in the form
\begin{equation}
    \mathcal{S}(n, a, b) = \langle Z_{i} X_{i \oplus a}  X_{i \oplus a \oplus b} Z_{i \oplus 2a \oplus b}| \forall i \in \mathbb{Z} / n \mathbb{Z} \rangle,
    \label{eq:def_cyclic_codes}
\end{equation}
where $n$ is the total number of qubits, $a$ and $b$ are positive integers, and $\oplus$ denotes addition modulo $n$. Each weight-four stabilizer generator is of XZZX type, with $a - 1$ identities inserted between $Z$ and $X$ and $b - 1$ identities inserted between two $X$s. 
We refer to the XZZX cyclic code defined in Eq.~\eqref{eq:def_cyclic_codes} as $\mathcal{S}(n, a, b)$. We note that 
$\mathcal{S}(7, 1, 3)$
has been considered in Ref.~\cite{robertson2017tailored} and shown to have a good performance against $Z$-biased noise.

For $Z$-biased noise, we can introduce a parameter $\eta = p_Z/(p_X + p_Y)$, which ranges from $\frac{1}{2}$ to infinity, to characterize the noise bias. 
We aim to find codes that are efficient over a wide range of biases $\eta$.
We attempt this by finding codes that can reach large effective code distance in the two extreme cases --- under depolarizing noise ($\eta = \frac{1}{2}$) and pure Pauli $Z$ noise ($\eta = \infty$) --- using only a small number of qubits. 
We may directly generalize the five-qubit code by keeping $a = b = 1$ and increasing $n$. 
However, in this way $d_Z$ increases while $d$ is fixed, e.g.\ $\mathcal{S}(13, 1, 1)$ has $d_Z = 13$ and $d = 3$. 
It turns out that to simultaneously increase $d_Z$ and $d$ we need to also modify the stabilizer structure, i.e.\ to change $a, b$. 
For example, the $\mathcal{S}(13, 2, 1)$ code has $d_Z = 13$ and $d = 5$. 
In general, it is easy to identify codes that have the maximal effective distance $d_Z = n$ (using $n$ qubits) under pure Pauli $Z$ noise since any code defined in Eq.~\eqref{eq:def_cyclic_codes} with $b$ and $n$ being coprime has a repetition-code structure by neglecting the $Z$ components in the stabilizers. 
However, it is nontrivial to identify codes that are also efficient against depolarizing or finite-bias noise. 
To accomplish this, we can wrap the cyclic codes on a torus and map them to a family of generalized toric codes (GTCs) introduced in Ref.~\cite{kovalev2012improved}.

~\\
\noindent
\textbf{From the XZZX cyclic codes to the XZZX generalized toric codes}\\
An XZZX generalized toric code $\textrm{GTC}(\vec{L}_1, \vec{L}_2)$ is a stabilizer code with qubits on a square lattice $\mathbb{Z}^2$ and stabilizers generators $\{S_{i,j} \equiv X_{i,j} Z_{i+1,j} Z_{i,j+1} X_{i + 1, j + 1}| i,j\in\mathbb{Z}\}$, with boundary conditions specified by the two basis periodicity vectors $\vec{L}_1, \vec{L}_2 \in \mathbb{Z}^2$: two points 
$\vec u,\vec v \in\mathbb{Z}^2$ 
are identified iff:
\begin{equation}
    \vec u-\vec v
    \in \textrm{span}(\vec{L}_1, \vec{L}_2) := \{m_{1} \vec{L}_{1}+m_{2} \vec{L}_{2} | m_{1}, m_{2} \in \mathbb{Z}\}.
    \label{eq:boundary_condition}
\end{equation}
$\textrm{GTC}(\vec{L}_1,\vec{L}_2)$ encodes $k$ logical qubits in $n$ physical qubits where $n = |\vec{L}_1 \times \vec{L}_2|$. 
If both periodicity vectors have even 1-norm, then $k = 2$. 
Otherwise, $k=1$ \cite{sarkar2021graph}.

A GTC can be viewed as stabilizer codes defined on a graph $G(\vec{L}_1, \vec{L}_2)$ embedded on a torus~\footnote{An embedding of a graph $G(V, E)$ with vertices $V$ and edges $E$ on a manifold $\mathcal{M}$ is a map $\Gamma: V \cup E \rightarrow \mathcal{M}$. With the embedding, we can define the plaquettes (or faces) of the embedded graph as $F = \mathcal{M} \backslash \Gamma(E)$. See Ref.~\cite{sarkar2021graph} for details. We refer to the graphs associated with the GTCs embedded graphs (on the torus) with well-defined vertices, edges and plaquettes, unless specially noted.}, with qubits on vertices and stabilizers on plaquettes. 
The infinite square lattice $\mathbb{Z}^2$ acts as the universal cover of $G(\vec{L}_1, \vec{L}_2)$, with the covering map given by the boundary condition Eq.~\eqref{eq:boundary_condition}. 
A code is uniquely specified by the submodule of $\mathbb{Z}^2$ given by $\textrm{span}(\vec{L}_1, \vec{L}_2)$, the span of the two basis vectors $\vec{L}_1, \vec{L}_2$. 
Different choices of basis periodicity vectors give the same GTC so long as they are related by a unimodular transformation. 
A single Pauli $Z_{i,j}$ ($X_{i,j}$) anti commutes with two stabilizer generators that lie along the diagonal: $\{ S_{i - 1,j - 1}, S_{i, j}\}$ ($\{ S_{i - 1,j}, S_{i, j - 1}\}$). 
To facilitate the analysis, we define the diagonal axes, which we call the ``XZ'' axes, to be the axes corresponding to the $X$ and $Z$ error chains with respective basis vectors $\hat{x} := (-1, 1), \hat{z} := (1,1)$. 
We note that the GTCs encoding two logical qubits, which can be obtained from the CSS toric codes~\cite{kitaev2003fault} by applying local Hadamard transformations and twisting the boundary conditions, are considered for biased noise in Ref.~\cite{roffe2022bias}. 
In this work, however, we will focus on the GTCs encoding 1 logical qubit since they can reduce the required code size by roughly a factor of 2 for reaching a target effective code distance compared to their 2-logical-qubit counterparts (which will become clear later). 
The rectangular-lattice toric codes considered in Ref.~\cite{ataides2021xzzx} with $\vec{L}_1 = (d-1, 0), \vec{L}_2 = (0, d)$ belong to the GTCs encoding 1 logical qubit. 
However, Ref.~\cite{ataides2021xzzx} only considers this special instance and has a limited discussion on its performance when the bias is finite. 
In this work, we will systematically investigate the performance of the 1-logical-qubit GTCs by studying their effective distance and adaptively find the optimal codes given any noise bias. 

The XZZX cyclic codes can be mapped to a subset of GTCs by wrapping the qubits around the torus along a certain direction. 
As an example, we show how the $\mathcal{S}(13, 2, 1)$ code can be mapped to the GTC with $\vec{L}_1 = (3,2), \vec{L}_2 = (-2,3)$ in Fig.~\ref{fig:cyclic_codes_to_GTCs}. 
We explicitly provide more general mappings from the XZZX cyclic codes to the GTCs in Supplemental Material \cite{SM}. 
We note that the GTCs are a larger family of codes that also include non-cyclic codes. 
By mapping the cyclic codes to the GTCs, we benefit from the following two aspects: 
(1) We can use topological tools to efficiently obtain the effective code distance given a finite noise bias, which enables us to adaptively design the optimal codes; 
(2) We can design efficient decoders that can lead to similarly high thresholds as those for the XZZX surface codes \cite{ataides2021xzzx}.

\begin{figure}[ht!]
    \centering
    \includegraphics[width = 0.48\textwidth]{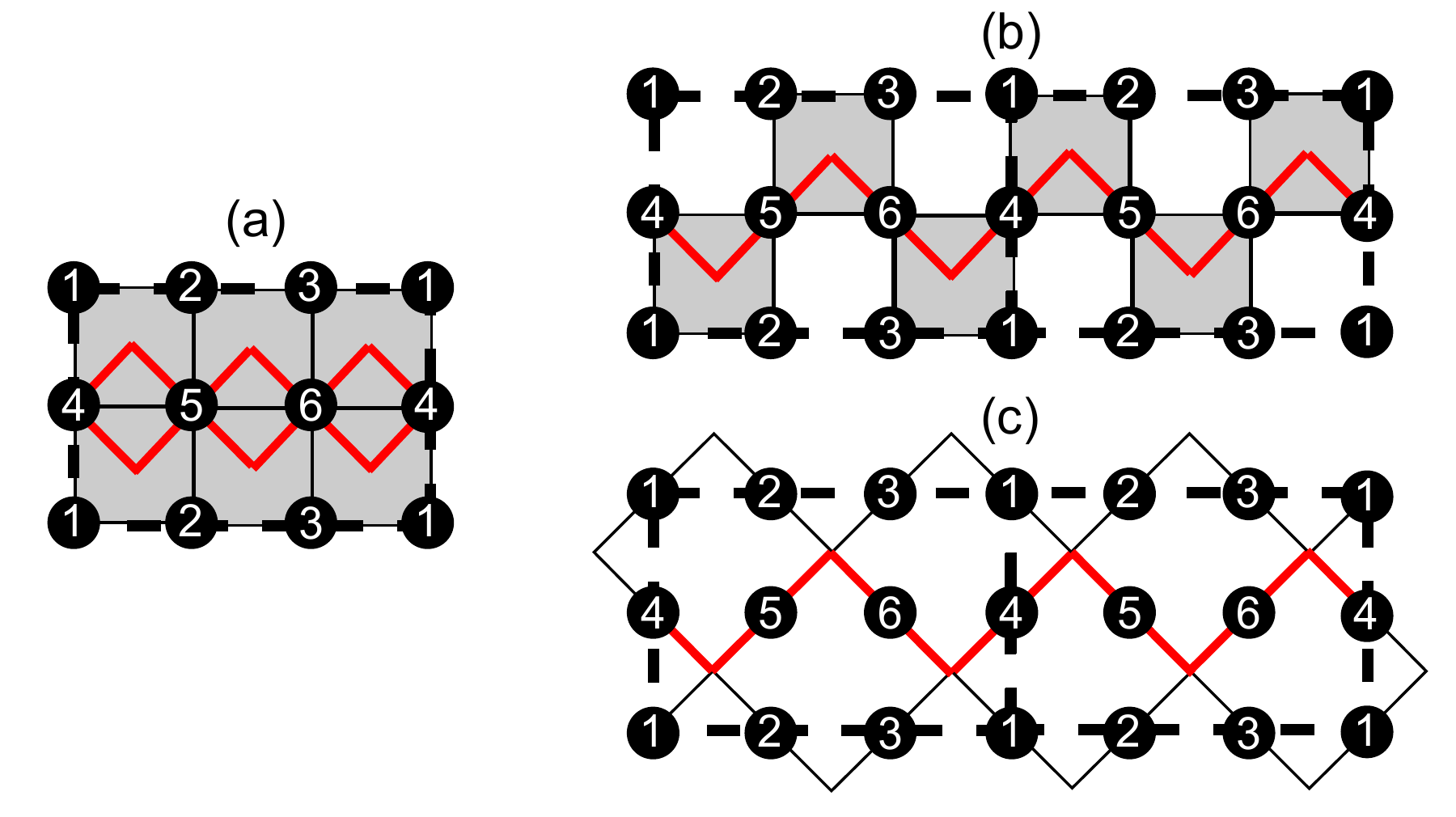}
     \caption{The construction of doubled graph for a non-two-colorable GTC with $\vec{L}_1 = (3,0), \vec{L}_2 = (0,2)$. 
     (a) The original non-two-colorable graph $G$ that defines the GTC. There is a XZZX stabilizer on each of the shaded plaquettes.
     The red cycle depicts a logical operator $Y_4 Y_5 Y_6$ that cannot wrap back on itself after a single loop due to odd periodicity in the horizontal direction. 
     (b) The doubled graph $G_d$ obtained by taking two copies of $G$ and gluing them horizontally. 
     Now $G_d$ becomes two colorable and the stabilizers are only put on the shaded plaquettes. 
     Consequently, a single loop is sufficient for the logical operator to wrap back on itself. 
     (c) The equivalent graph $G^{\prime}_d$ of $G_d$ where qubits are placed on the edges and stabilizers are placed on the vertices. 
     Now the logical operator corresponds to a well-defined cycle that is homologically nontrivial on the doubled torus. 
    }
    \label{fig:Doubling}
\end{figure}

~\\
\noindent
\textbf{Deriving the effective code distance for the GTCs}\\
Calculating the effective code distance is likely to be computationally intractable in general since even computing the distance of a classical linear code is NP-hard. 
However, logical operators of topological codes embedded in a manifold are easily identified with geometrical objects on the manifold, and therefore the effective code distance of the GTCs can be efficiently derived using topological tools.
Recall that a $\textrm{GTC}(\vec{L}_1, \vec{L}_2)$ is defined on an embedded graph $G(\vec{L}_1, \vec{L}_2)$ on a torus. 
We first consider the case when the plaquettes of $G$ are two-colorable, i.e.\ one can consistently two color the plaquettes such that two plaquettes sharing the same edge have different colors. 
In this case, we can transform $G$ to a graph $G^{\prime}$ in which qubits are associated with edges while stabilizers are associated with plaquettes and vertices (e.g.\ from Fig.~\ref{fig:Doubling}(b) to Fig.~\ref{fig:Doubling}(c)). 
Now $G^{\prime}$ is the same as the Kitaev's construction \cite{kitaev2003fault} and mathematically $G$ is the medial graph of $G^{\prime}$. 
As a result, the GTCs associated with $G^{\prime}$ can be described by the standard 2 chain complex for CSS toric codes \cite{bravyi2014homological}, and the logical operators are associated with homologically nontrivial cycles on the torus. 
See Supplementary Material~\cite{SM} for more details. 
In fact, the two-colorable GTCs are equivalent to the CSS toric codes by local Hadamard transformation. 
The distance of these two-colorable codes can then be obtained by estimating the shortest length of the nontrivial cycles, and the effective distance under an asymmetric noise, as we will show later, equals the shortest length of the nontrivial cycles under a noise-modified distance metrics.

However, the GTCs of interest that encode 1 logical qubit, are defined by embedded graphs $G$ that are not two-colorable, when at least one of $\vec{L}_1, \vec{L}_2$ is odd in 1-norm. 
In this scenario, the graph $G$ cannot be consistently two-colored and as a consequence, it can not be directly transformed to a graph $G^{\prime}$ corresponding to a 2 chain complex.
Fortunately, we can still use the algebraic tools by constructing the ``doubled" graph $G_d(\vec{L}_{1,d}, \vec{L}_{2,d})$ \cite{sarkar2021graph}: 
Without loss of generality, we assume $\vec{L}_1$ is even while $\vec{L}_2$ is odd in 1-norm. 
We then obtain the doubled graph $G_d$ by combining two copies of $G$ together along $\vec{L}_1$. 
Topologically, this corresponds to taking two tori, cutting them open along the $\vec{L}_1$ cycle and gluing them together. 
As such, $G_d$ is embedded on the doubled torus, which is specified by two doubled periodicity vectors $\vec{L}_{1,d}, \vec{L}_{2,d}$ (similar as that the original torus is specified by the periodicity vectors $\vec{L}_{1}, \vec{L}_2$ via Eq.~\eqref{eq:boundary_condition}) that are given by the following map: $\vec{L}_{1,d} = \vec{L}_1, \vec{L}_{2,d} = 2 \vec{L}_2$.
As an example, we show how we construct the doubled graph for a GTC with $\vec{L}_1 = (0,2), \vec{L}_2 = (3,0)$ in Fig.~\ref{fig:Doubling} (from (a) to (b)). 

The doubled graph $G_d$ is now two colorable. 
We can then transform $G_d$ to a graph $G_d^{\prime}$ with the original vertices in $G_d$ on the edges of $G_d^{\prime}$. 
The black plaquettes in $G_d$ are transformed into vertices in $G_d^{\prime}$. 
See the transformation from Fig.~\ref{fig:Doubling}(b) to (c). 
We now associate the edges in $G_{d}^{\prime}$ with Pauli operators up to phases and vertices/plaquettes with stabilizer generators. 
The idea of the doubling was introduced and analyzed in pure graph-based formalism in Ref.~\cite{sarkar2021graph}. 
Here we formalize the codes associated with the doubled graph from the perspective of algebraic topology. 
As detailed in Methods, the representatives of logical operators $Z_L, X_L, Y_L$ are associated with the nontrivial elements in the first homology group, or equivalently, three homologically nontrivial loops, that are defined on the \textit{doubled} torus. 
We note that because of the doubling, the non-two-colorable GTCs can effectively reduce the code size required for reaching a certain effective distance by a factor of 2 compared to their two-colorable counterparts, thus being more resource efficient.

With the above topological construction, we can then use geometrical method to calculate the effective code distance $d^{\prime}$. 
This can be readily calculated when $X$ and $Z$ errors are independent. 
Therefore, we first consider independent Pauli X and Z noise, in which the probability distribution is given by $p_X = p_Z^{\omega}$, $p_Y = p_X p_Z = p_Z^{\omega + 1}$ and $p = p_Z + p_X + p_Y$, where we assume $\omega \geq 1$.
We note that here we use a bias parameter $\omega$ that is different from the parameter $\eta \equiv p_Z/(p_X + p_Y)$ commonly used in the literature \cite{tuckett2018ultrahigh, tuckett2019tailoring, tuckett2020fault, ataides2021xzzx}. 
For independent Pauli $X$ and $Z$ noise, we can convert $\eta$ to $\omega$ by $\omega = \frac{\log \eta}{\log 1/p_Z} + \frac{\log (1 + 1/p_Z)}{\log 1/p_Z}$, where $\omega$ depends on both $\eta$ and the error probability $p_Z$. 
Under such a noise model, the modified weights of Paulis are
$\textrm{wt}'(Z)= 1$, $\textrm{wt}'(X)=\omega$ and $\textrm{wt}'(Y) = \omega + 1$.
Then, the effective code distance is given by the length of shortest homologically nontrivial cycle on the doubled torus with distance metrics being the rescaled 1-norm (in the XZ axes):
\begin{equation}
\begin{array}{c}
    d^{\prime}=\min _{m_{1}, m_{2} \in \mathbb{Z}}\left\|m_{1} \vec{L}_{1, d}+m_{2} \vec{L}_{2, d}\right\|_{xz, 1}^{\prime}, \\
\end{array}
\label{eq:effective_code_distance}
\end{equation}
where the rescaled 1-norm of a vector $\alpha \hat{x} + \beta \hat{z}$ is given by $\|\alpha \hat{x} + \beta \hat{z}\|_{xz,1}^{\prime}\equiv\omega|\alpha|+|\beta|$.
In Methods we present an efficient algorithm with complexity $O(d^{\prime^{2}})$ to compute the effective distance $d^{\prime}$.

\begin{figure}[ht!]
    \centering
    \includegraphics[width = 0.48\textwidth]{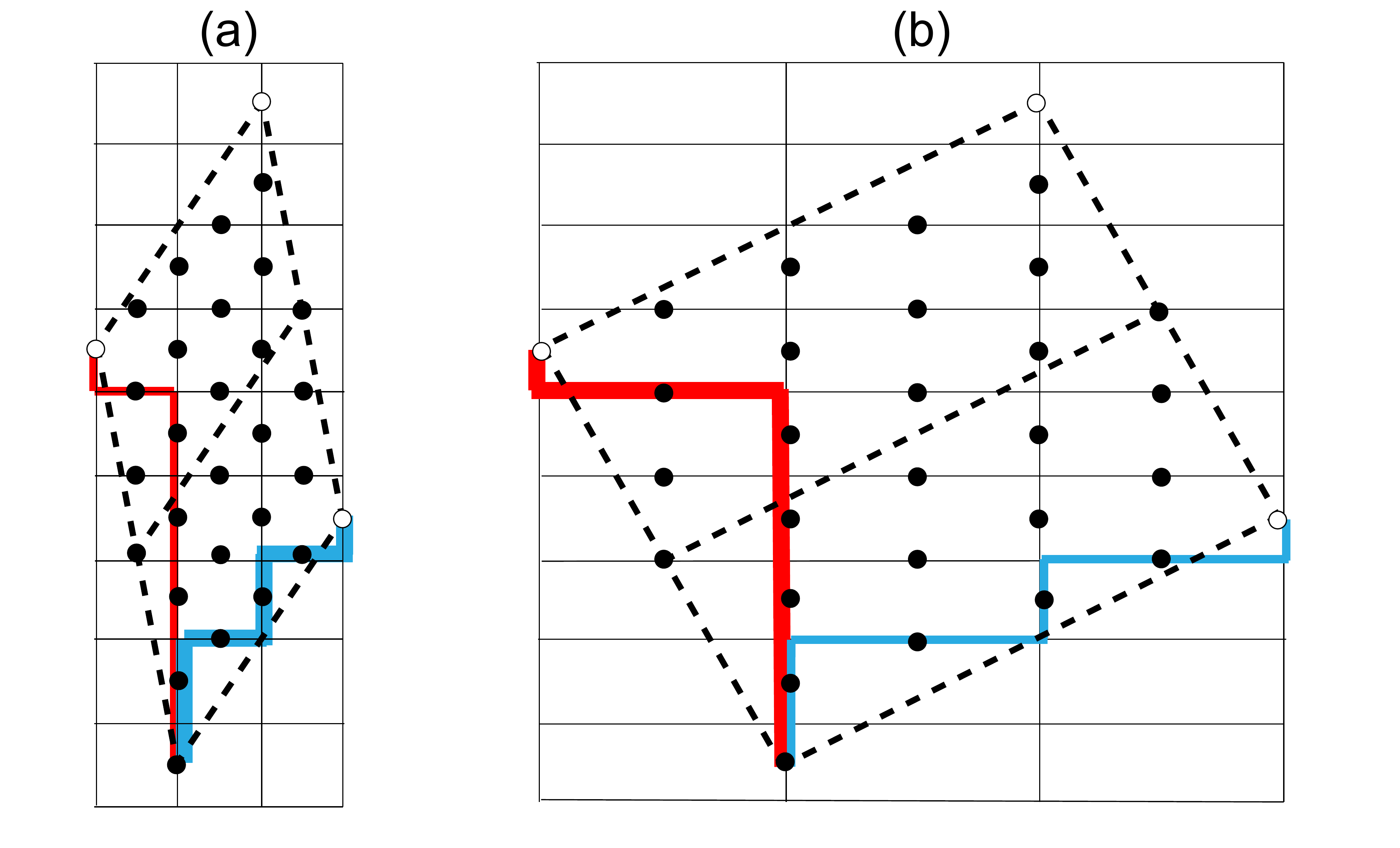}
     \caption{Geometric representation of the logical operators of the $[[13, 1, 5]]$ GTC with $\vec{L}_1 = (-1,5) = 2\hat{x} + 3\hat{z}$, $\vec{L}_2 = (-3,2) = -\frac{1}{2}\hat{x} + \frac{5}{2}\hat{z}$ for (a) $\omega = 1$ and (b) $\omega = 3$. 
     The horizontal and vertical axes are aligned with the $X$ and $Z$ axes, respectively. 
     The doubled graph is obtained by doubling along $\vec{L}_2$: $\vec{L}_{1,d} = \vec{L}_1, \vec{L}_{2,d} = 2 \vec{L}_2 = - \hat{x} + 5\hat{z}$. 
     The blue (red) cycle represents the logical operator associated with $\vec{L}_{1,d}$ ($\vec{L}_{2,d}$). 
     The shortest nontrivial cycles $\vec{L}_{m}$ in the modified 1-norm determined by Eq.~\eqref{eq:effective_code_distance} are thickened. 
     For (a) $\omega = 1$, $\vec{L}_m = \vec{L}_{1,d}$ and $d^{\prime} = \|\vec{L}_{1,d}\|_{xz,1} = 5$; For (b) $\omega = 3$, $\vec{L}_m = \vec{L}_{2,d}$ and $d^{\prime} = \|\vec{L}_{2,d}\|_{xz,1} = 8$.}
    \label{fig:geometric_rep_logical_operators}
\end{figure}

It is worth noting that the choice of the shortest nontrivial cycle, which corresponds to the logical operator with minimum effective weight, depends on the noise bias $\omega$. 
Moreover, the effective distance $d^{\prime}$ for a given GTC also varies with $\omega$ and typically increases with $\omega$. 
As an example, we show how we can geometrically find the minimum-effective-weight logical operators and obtain the effective code distance for the [[13,1,5]] GTC under different bias in Fig.~\ref{fig:geometric_rep_logical_operators}. 

\begin{figure}[ht!]
    \centering
    \includegraphics[width = 0.48\textwidth]{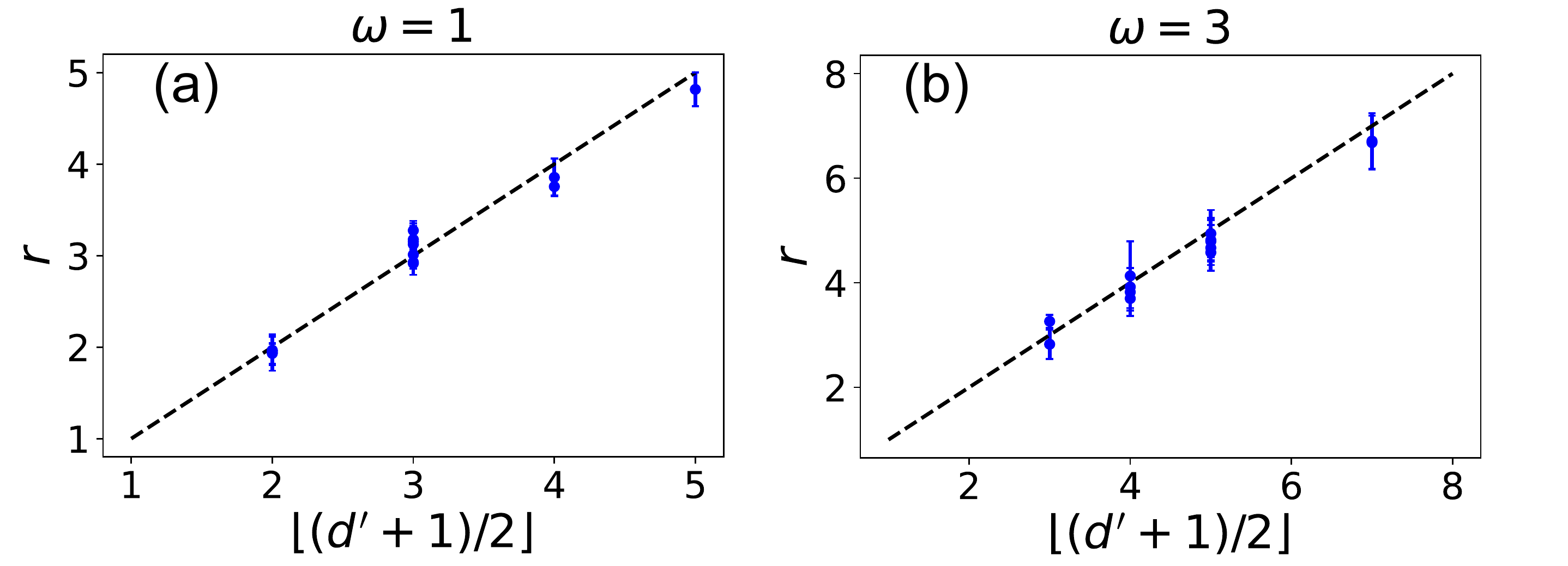}
     \caption{Verification of the effective code distance Eq.~\eqref{eq:effective_code_distance} as a good performance metrics for the GTCs. 
     Given a bias parameter $\omega$, we numerically obtain the logical error rate $p_L$ for a set of randomly chosen GTCs using the MWPM decoders over a range of physical error rate $p$ (below the threshold), and fit the logical error rate by $p_L \propto p^r$. 
     There is a good agreement between $r$ and $\lfloor{(d^{\prime} + 1)/2}\rfloor$ for both (a) $\omega = 1$ and (b) $\omega = 3$. 
     The range of the physical error rate $p$ used for the fitting is: (a) $p \in [0.02, 0.08]$ and (b) $p \in [0.06, 0.1]$.}
    \label{fig:verification_of_effective_code_distance}
\end{figure}

To verify that the effective code distance estimated via Eq.~\eqref{eq:effective_code_distance} is indeed a good proxy for the code performance under the independent XZ noise model, we perform the Monte Carlo (MC) simulation using a MWPM decoder (see Methods) and fit the logical error rate by $p_L \propto p^r$. 
We compare the numerically fitted exponent $r$ with $\left\lfloor\left(d^{\prime}+1\right) / 2\right \rfloor$, where $\left\lfloor \cdot \right \rfloor$ represents the floor operation, in Fig.~\ref{fig:verification_of_effective_code_distance}.

We can see that for most of the codes $r$ agrees well with $\left\lfloor\left(d^{\prime}+1\right) / 2\right\rfloor$ within the numerical uncertainty. 
The systematical deviation for some codes under $\omega = 3$ occurs because when $\omega > 1$, different Pauli operators have different effective weights and the effective weight of the most probable uncorrectable error associated with a logical operator with effective weight $w^{\prime}$ is not necessarily (and in fact, only lowered bounded by) $\left\lfloor\left(w^{\prime}+1\right) / 2\right\rfloor$. 
As a result, $r$ is, in general, only lower bounded by $\left\lfloor\left(d^{\prime}+1\right) / 2\right\rfloor$. 
In Methods, we provide an improved approximation for $r$.

~\\
\noindent
\textbf{Code thresholds}\\
The topological construction of the GTCs indicates that the GTCs could potentially have exceptionally high thresholds, which might be further boosted by having large bias.  
First we note that the GTCs are locally equivalent to the XZZX surface codes and differ mainly by boundary conditions. 
As a result, the code-capacity threshold, an asymptotic quantity for asymptotically-large code blocks, is the same for two code families if optimal decoders are applied. 
Next, we show that the GTCs have similarly high thresholds using our tailored efficient MWPM decoders. 
\begin{figure}[ht!]
    \centering
    \includegraphics[width = 0.5\textwidth]{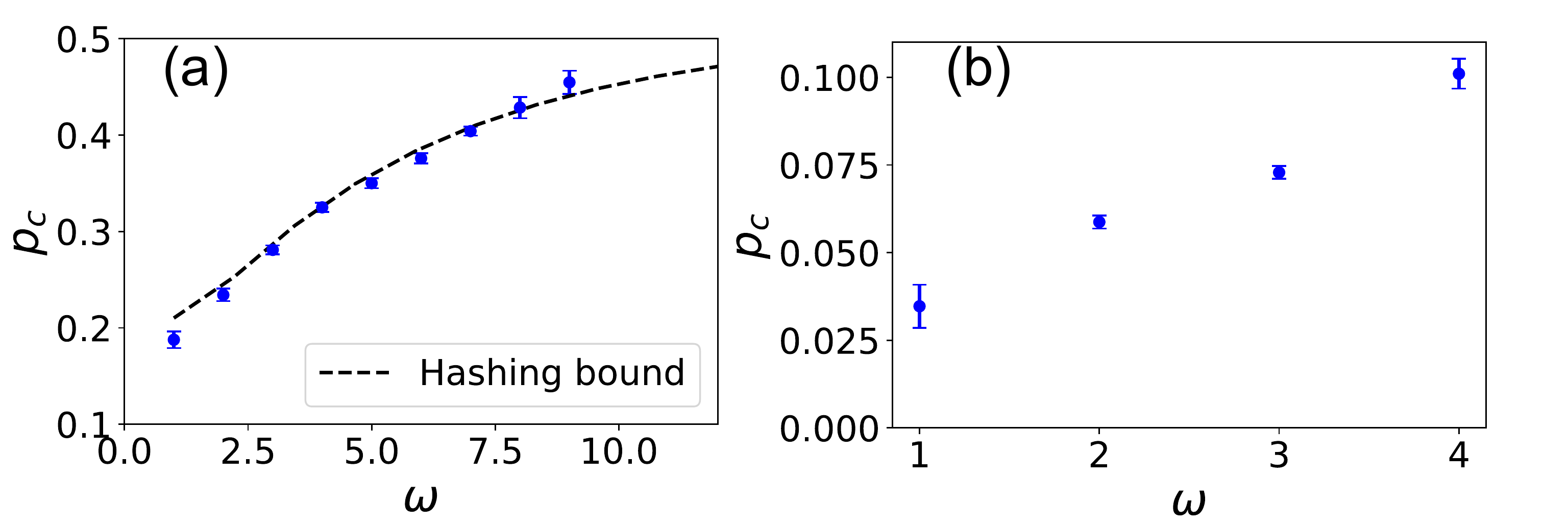}
     \caption{The thresholds $p_c$ of GTCs using MWPW decoders as functions of the bias parameter $\omega$, obtained by doing the critical exponent fit \cite{wang2003confinement} on numerical data from MC simulations. 
     (a) Code threshold assuming perfect syndrome measurements. The dashed line indicates the hashing bound. 
     (b) Code threshold under a phenomenological error model, in which each syndrome measurement fails with a probability that equals the total error probability $p$ of the data qubits.}
    \label{fig:threshold}
\end{figure}

We adopt the independent $X$ and $Z$ noise model and numerically extract the thresholds $p_c$ of GTCs using MWPW decoders for different bias parameter $\omega$ by doing the critical exponent fit \cite{wang2003confinement} on the numerical data from the MC simulations. 
In Fig.~\ref{fig:threshold}(a), we plot the thresholds of the GTCs when assuming perfect syndrome extractions. 
The thresholds match or even surpass the hashing bound (indicated by the dashed line), similarly as the XZZX surface codes. 
In Fig.~\ref{fig:threshold}(b), we plot the thresholds under a phenomenological noise model, in which each syndrome measurement fails with a probability that equals to the total error probability $p$ of the data qubits. 
The phenomonological thresholds increase from $3.5\%$ to $10\%$ as $\omega$ increases from 1 to 4. 

\noindent
~\\
\textbf{Adaptive code design for qubit-efficient QEC} \\
Under infinitely $Z$-biased noise, the optimal GTCs (encoding one logical qubit)  should correspond to the cyclic codes with a repetition structure, whose effective distance reaches the optimal linear scaling with $n$, i.e.\ $d_{Z} = n$. 
Here we explicitly identify these cyclic GTCs. 
A GTC is cyclic if there is a cycle of length $n$ along certain direction $\hat{l}_0 \in \mathbb{Z}^2$ with $a,b$ being coprime, on the torus that goes through all the qubits without repetition, e.g.\ the grey string in Fig.~\ref{fig:cyclic_codes_to_GTCs}. 
The qubits are then labeled along this cycle to be a XZZX cyclic code. 
Given such a direction $\hat{l}_0$, we say that the GTC is cyclic along $\hat{l}_0$. 
We then identify the GTCs that correspond to the XZZX cyclic codes with repetition structures by identifying the direction along which they are cyclic: A GTC has a repetition-Z structure iff it is cyclic along (1,1) direction. 
We prove this structure and show that the GTCs can also have repetition structures for the pure $X$ and $Y$ noise (with $d_X = n$ and $d_Y = n$, respectively) in Methods.

In the finite-bias regime, we can use our topological construction and geometrical methods to adaptively identify the optimal GTCs given any bias parameter $\omega$. 
We restrict ourselves to non-two-colorable GTCs since they are more resource efficient. 
The task is now to find the GTC that uses the smallest number of physical qubits $n$ to reach a given effective code distance $d^{\prime}$. 
Given the geometrical interpretation of $d^{\prime}$ and $n$: $d^{\prime} = \min _{m_{1}, m_{2} \in \mathbb{Z}}\left\|m_{1} \vec{L}_{1, d}+m_{2} \vec{L}_{2, d}\right\|_{xz,1}^{\prime}$ and $n = \tfrac{1}{2}\left|\vec{L}_{1, d} \times \vec{L}_{2, d}\right|$,
this task is equivalent to finding the densest packing of diamonds whose aspect ratio is given by the bias parameter $\omega$. 

\begin{figure}[ht!]
    \centering
    \includegraphics[width = 0.5\textwidth]{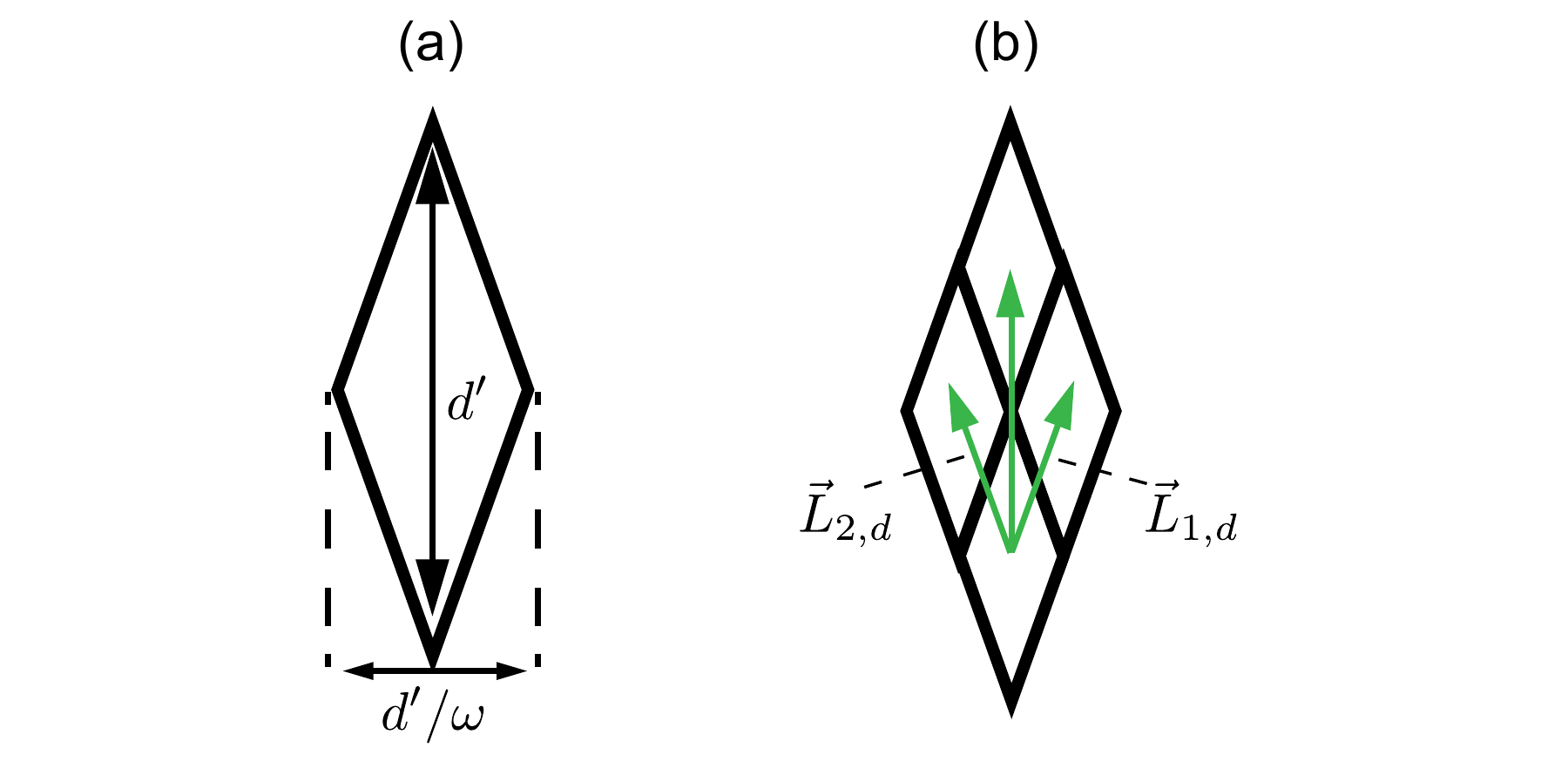}
     \caption{The densest packing of diamonds that correspond to the optimal choice of the GTCs given a bias parameter $\omega$. 
     The diamonds to pack (a) have aspect ration $\omega$ and the densest packing pattern is the regular tiling (b). }
    \label{fig:densest_packing}
\end{figure}

\begin{figure}[ht!]
    \centering
    \includegraphics[width = 0.49\textwidth]{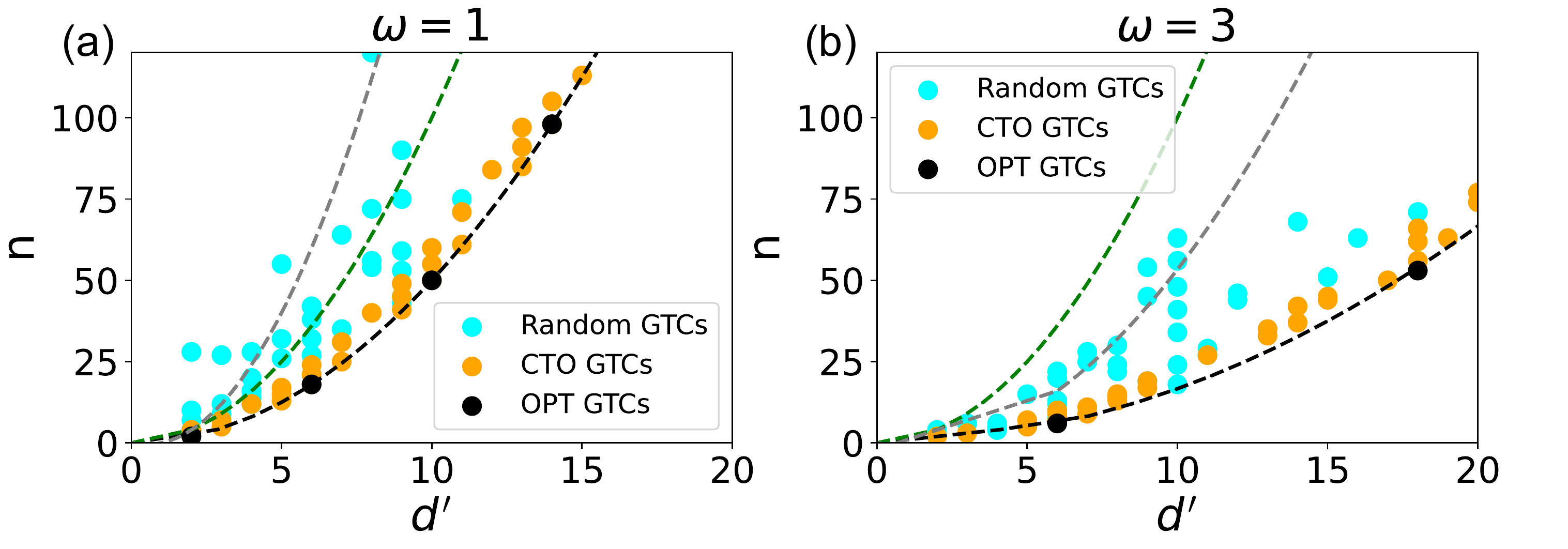}
     \caption{The scatter plots of the required size $n$ of different codes for achieving a target effective distance $d^{\prime}$ under (a) $\omega = 1$, (b) $\omega = 3$. The black dashed line indicates the bound (Eq.~\eqref{eq:code_upper_bound}) $n = d^{\prime 2}/2\omega$ for the GTCs. 
     The green dashed line indicates the standard scaling $n = d^{\prime 2}$ for $d^{\prime}$ by $d^{\prime}$ rotated planar surface codes. 
     The grey dashed line indicates the scaling $n = \max [2d^{\prime 2}/\omega - d^{\prime}(1 + 1/\omega), 3d^{\prime} - 2]$ for the unrotated planar surface codes with optimized aspect ratio. 
     All the codes are of the XZZX type.}
    \label{fig:code_histogram}
\end{figure}

As shown in Fig.~\ref{fig:densest_packing}(a), the diamond to pack has diagonals with length $d^{\prime}$ and $d^{\prime}/\omega$, respectively. 
Obviously, the densest packing pattern is the regular tiling of a surface using the diamonds (see Fig.~\ref{fig:densest_packing}(b)). 
Therefore, the optimal choice of $\vec{L}_{1, d}, \vec{L}_{2, d}$ is:
\begin{equation}
        \vec{L}_{1, d}^{\textrm{OPT}}=\tfrac{1}{2} d^{\prime}\left(\tfrac{1}{\omega} \hat{x}+\hat{z}\right),\quad \vec{L}_{2, d}^{\textrm{OPT}}=\tfrac{1}{2} d^{\prime}\left(- \tfrac{1}{\omega} \hat{x}+\hat{z}\right).
    \label{eq:optimal_choice_Ls}
\end{equation}
Eq.~\eqref{eq:optimal_choice_Ls} gives $n=d^{\prime 2} / 2 \omega$, or $d^{\prime} = \sqrt{2 \omega n}$. 
However, the densest packing pattern is not always achievable since the codes have an additional constraint that $d^{\prime} \leq n$, as there is always a logical operator consisting of all Pauli Zs of effective weight $n$.
Hence, the optimal codes satisfy:
\begin{equation}
n= \begin{cases}d^{\prime} & d^{\prime} \leq 2 \omega \\ d^{\prime 2}/2\omega & d^{\prime}>2 \omega\end{cases},
    \label{eq:code_upper_bound}
\end{equation}
or equivalently, $d^{\prime} = \min (n , \sqrt{2\omega n})$. Eq.~\eqref{eq:code_upper_bound} should be viewed as an (tight) lower bound on the required size of our XZZX codes family for reaching a target effective distance (see the black dashed curve in Fig.~\ref{fig:code_histogram}). 
Furthermore, Eq.~\eqref{eq:code_upper_bound} provides the guiding principles for choosing the optimal codes: Given a noise bias $\omega$ and a target effective distance $d^{\prime}$, the cyclic GTCs with a repetition-code structure are optimal for $d^{\prime} \leq 2\omega$; While the GTCs with an optimized layout (optimal choice of $\vec{L}_1, \vec{L}_2$), which corresponds to the densest packing pattern of the associated diamonds, are optimal for $d^{\prime} > 2\omega$.

However, due to the constraints that the periodicity vectors $\vec{L}_1, \vec{L}_2$ should be in $\mathbb{Z}^2$ and they produce a non-two-colorable graph, the optimal codes that saturate the upper bound are sparse in both $d^{\prime}$ and $n$ (see the black dots in Fig.~\ref{fig:code_histogram}). 
To obtain more codes with good performance, we define the following close-to-optimal (CTO) codes by finding the codes that correspond to close-to-optimal packing patterns:
\begin{equation}
\vec{L}_{i, d}^{\mathrm{CTO}}=\vec{L}_{i, d}^{\mathrm{OPT}}+\delta l_{i}, \quad\left\|\delta l_{i}\right\|_{x z, 1} \leq \Delta,
\label{eq:close-to-optimal_code_choice}
\end{equation}
for $i = 1,2$. 
Here $\Delta$ is a parameter that characterizes how far the packing pattern given by $\vec{L}_{i, d}^{\textrm{CTO}}$ is deviated from the densest packing pattern (given by $\vec{L}_{i, d}^{\textrm{OPT}}$).
By increasing $\Delta$ we further relax the packing pattern and obtain more CTO codes. 

In Fig.~\ref{fig:code_histogram}, we plot the required size $n$ of different codes as a function of the target effective distance $d^{\prime}$ for $\omega = 1$ and 3. 
We compare the tailored GTCs with the unrotated planar XZZX surface code \cite{ataides2021xzzx}, whose aspect ratio is optimized according to $\omega$. For $d^{\prime} \leq 2\omega$, the tailored GTCs enjoys the optimal scaling between $n$ and $d^{\prime}$, i.e. $n = d^{\prime}$, because some GTCs can have a repetition-code structure while the planar codes cannot;
For $d^{\prime} > 2\omega$, $n$ scales quadratically with $d^{\prime}$ for both code families, but the planar codes require roughly four times more qubits than the tailored GTCs.

\begin{figure}[ht!]
    \centering
    \includegraphics[width = 0.5 \textwidth]{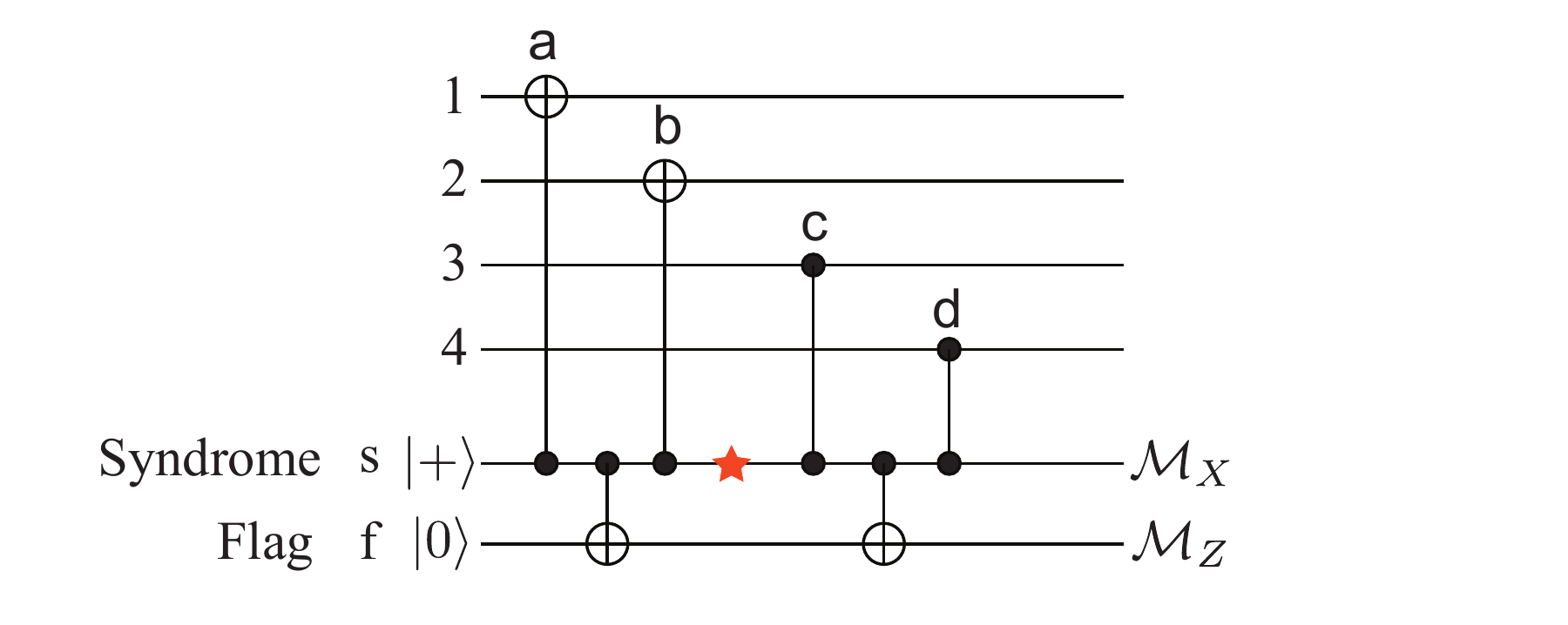}
     \caption{The flag circuit to extract a weight-4 stabilizer $X_1 X_2 Z_3 Z_4$. 
     We prepare a syndrome qubit s in $|+\rangle$ state, apply a sequence of $CX$ (a,b) and $CZ$ (c,d) gates between the s and the data qubits, and measure s in the $X$ basis to obtain the syndrome associated with $X_1 X_2 Z_3 Z_4$. 
     In addition, we apply two $CX$ gates between s and an extra flag qubit f to catch the bad (bit-flip) errors on the syndrome qubit. 
     Without the flag qubit the bad errors after gates $b$ and $c$ can propagate to the data qubits with larger (effective) weight. 
     With an extra flag qubit, the bad errors (e.g.\ a Pauli $X$ error depicted by the red star) can be detected and corrected adaptively. }
    \label{fig:flag_circuit}
\end{figure}

~\\
\noindent
\textbf{Fault-tolerant QEC}\\
In this section we present a fault-tolerant QEC scheme for the GTCs using flag qubits. 
The main idea of the flag fault tolerance \cite{chao2018quantum, chao2018fault, chamberland2018flag, reichardt2020fault} is that a small number of extra qubits (flags) are used to catch the bad ancilla errors that propagate to higher-weight data errors during the stabilizer measurements (e.g.\ a Pauli $X$ error on the syndrome qubit s that occurs after gate b, which is depicted by the red star, during the stabilizer measurement shown in Fig.~\ref{fig:flag_circuit}). 
Therefore, the code distance is preserved under a circuit-level noise, i.e.\ a distance-$d$ code can tolerate up to $t = \lfloor{(d-1)/2}\rfloor$ faults that occur at arbitrary locations during the protocol. 

When considering the biased noise, we have shown that it is the effective code distance, which is typically larger than distance for tailored codes, that characterizes the code performance. 
As such, we need to design FT schemes that preserve the effective code distance.
Following Refs.~\cite{aliferis2005quantum, gottesman2010introduction} we give the following definition of $t^{\prime}$-FTEC under biased noise: For $t^{\prime} = (d^{\prime} - 1)/2$, an error correcting protocol using a stabilizer code with effective code distance $d^{\prime}$ is $t^{\prime}$ fault-tolerant if the following two conditions are satisfied:

1. For an input code word with error of effective weight $s_1^{\prime}$, if any faults with effective weight $s_2^{\prime}$ occur during the protocol with $s_1^{\prime} + s_2^{\prime} \leq t^{\prime}$, ideally decoding the output state gives the same code word as ideally decoding the input state.

2. For faults with effective weight $s^{\prime}$ during the protocol with $s^{\prime} \leq t^{\prime}$, no matter how many errors are present in the input state, the output state differs from a code word by an error of at most effective weight $s^{\prime}$.

For the GTCs, we can use the flagged circuit in Fig.~\ref{fig:flag_circuit} to measure the XZZX stabilizers and achieve the above defined fault tolerance. 
More precisely, we claim that by measuring the stabilizers using the flagged circuits (and unflagged circuits) in an appropriate sequence and applying proper decoding, a GTC with effective distance $d^{\prime}$ can realize $t^{\prime}$-FTEC, where $t^{\prime} = (d^{\prime} - 1)/2$.

We prove the claim in Methods and only sketch it here.
For the flag-QEC scheme to work, two conditions have to be satisfied. 
(1) Errors with effective weight up to $t^{\prime}$ on the ancilla qubits that propagate to higher (effective) weight errors have to be detected by the flag qubits. 
(2) The bad errors sharing the same flag pattern have to be distinguishable and correctable by the Knill-Laflamme conditions. 
The first condition is guaranteed by using the flag circuit Fig.~\ref{fig:flag_circuit} while the satisfaction of the second condition is in general code-specific and is proved for the GTCs in Methods.

\section{Discussion}
\begin{figure}[ht!]
    \centering
    \includegraphics[width = 0.48\textwidth]{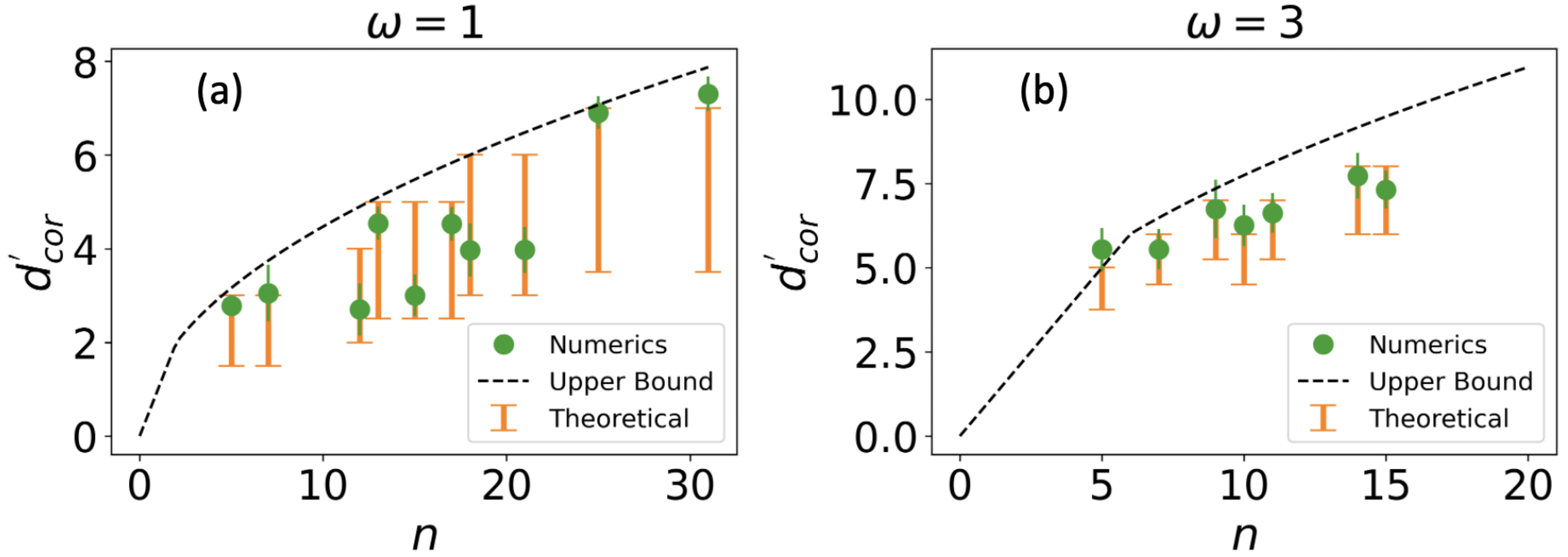}
     \caption{Performance of the close-to-optimal GTCs under the correlated Pauli $X$ and $Z$ noise for (a) $\omega = 1$ and (b) $\omega = 3$. 
     The dashed line indicates the bound Eq.~\eqref{eq:code_upper_bound}. 
     The orange bars indicate the effective code distance bounds given in Eq.~\eqref{eq:biased_effective_distance}. 
     The green dots are given by $2r - 1$, where $r$ is the exponent in the expression $p_L \propto {p^r}$ that is extracted from fitting the MC simulations using tensor network decoders. 
     The green bars indicate the standard numerical error. 
     The range of the physical error rate $p$ used for the fitting is: (a) $p \in [0.01, 0.03]$ and (b) $p \in [0.04, 0.07]$.}  
    \label{fig:biaseddist}
\end{figure}

\noindent
\textbf{Correlated Pauli $X$ and $Z$ noise}\\
In this section, we discuss the performance of the GTCs under another physically relevant noise model, which we call the correlated Pauli $X$ and $Z$ noise: $p_X = p_Y = p_Z^{\omega}$ and $p_X + p_Y + p_Z = p$, where $\omega \geq 1$.
This model does not assume the independence between $X$ and $Z$ errors and is widely considered in studying QEC under biased noise \cite{tuckett2018ultrahigh, tuckett2019tailoring, tuckett2020fault, ataides2021xzzx}. 
Under such a model, the effective weights of the Pauli operators are $\textrm{wt}'(Z) = 1$ and $\textrm{wt}'(X)=\textrm{wt}'(Y) =\omega$. 
It turns out that we can easily extend our analysis of the GTCs under the independent $X$ and $Z$ noise to the case where the $X$ and $Z$ error becomes correlated, and obtain similar results. 
Specifically, we have the following results:

(1) The effective code distance $d^{\prime}_{\textrm{cor}}$ of the GTCs under the correlated $X$ and $Z$ noise is close to that under the independent $X$ and $Z$ noise $d^{\prime}$ for the same $\omega$, especially in the high-bias regime. 
More precisely, we have
\begin{equation}
    \tfrac{\omega}{\omega + 1} d^{\prime} \leq d^{\prime}_{\textrm{cor}} \leq d^{\prime}.
    \label{eq:biased_effective_distance}
\end{equation}
Both the upper and lower bounds are tight and they converge to be the same as $\omega$ increases. 
Therefore, we can use Eq.~\eqref{eq:effective_code_distance} that we developed for calculating the $d^{\prime}$ to approximately estimate the $d^{\prime}_{\textrm{cor}}$ of the GTCs. 
In Fig.~\ref{fig:biaseddist}, we compare the bounds in Eq.~\eqref{eq:biased_effective_distance} (orange bars) with the numerical estimates of the $d^{\prime}_{\textrm{cor}}$ (green dots) for the close-to-optimal codes defined in Eq.~\eqref{eq:close-to-optimal_code_choice}. 
The numerical estimates of the $d^{\prime}_{\textrm{cor}}$ are given by $2r - 1$, where $r$ is the exponent in the expression $p_L \propto {p^r}$ that is extracted from fitting the MC simulations using tensor network decoders (see Methods for details of the decoders). 
The numerical estimates fall well within the theoretical bounds. 
For $\omega = 1$, the bounds Eq.~\eqref{eq:biased_effective_distance} are relatively loose due to the factor $\frac{\omega}{\omega + 1} = \frac{1}{2}$. 
As $\omega$ increases, the bounds become tighter and $d^{\prime}$ serves as a good approximation for $d^{\prime}_{\textrm{cor}}$ for all the codes.

(2) Based on (1), the optimal achievable performance of the GTCs under the biased model is close to Eq.~\eqref{eq:code_upper_bound}. 
The optimal/close-to-optimal codes under the correlated $X$ and $Z$ noise are among the close-to-optimal codes under the independent $X$ and $Z$ noise. 
To identify the former, we need to look for codes in the latter family with the minimal number of $Y$s in the shortest logical operators. 
Consequently, these codes have effective distance $d^{\prime}_{\textrm{cor}} \approx d^{\prime}$, which, according to the upper bound in Eq.~\eqref{eq:biased_effective_distance}, is the largest achievable effective distance (given a bias and code size). 
Such optimal/close-to-optimal codes under the correlated $X$ and $Z$ noise can be identified in Fig.~\ref{fig:biaseddist}, as those with the numerically extracted effective code distance (green dots) close to Eq.~\eqref{eq:code_upper_bound} (dashed line).

\noindent
\textbf{Practical applications}\\
In practice, the noise channel in many physical systems is asymmetric, e.g.\ noise biased towards dephasing. 
Here we consider the stabilized cat qubits in bosonic systems, whose bit-flip rate is exponentially suppressed by the cat size $|\alpha|^2$ while phase-flip rate is only linearly amplified by $|\alpha|^2$, thereby supporting exponentially large noise bias \cite{lescanne2020, grimm2020}. 
More importantly, the stabilized cats support a set of gates that preserve the bias of the noise \cite{puri2020bias, guillaud2019repetition, xu2021engineering}, which are important for fault-tolerant QEC in the circuit level. 
According to Refs.~\cite{chamberland2020building, darmawan2021practical, xu2021engineering}, it is expected that the error rates in these systems can reach $p_Z \sim 10^{-2}, p_X \sim 10^{-6}$, which corresponds to $\omega \approx 3$. 
Under this bias, the optimized GTCs can act effectively as repetition codes with $d^{\prime} = n$ for $n$ up to 6. 
For larger $n$, the GTCs remains resource efficient. 
As an example, the GTC with $\vec{L}_1 = (7,5), \vec{L}_2 = (-2, 1)$ has an  effective distance 9 using only 17 qubits.
In contrast, to achieve the same distance with the standard surface code under the depolarizing noise one would need 81 qubits. 
Furthermore, if we are not restricted by local connectivity, by adding only two extra qubits we can fault-tolerantly operate this GTC using only 19 physical qubits.

~\\
\noindent
\textbf{Outlook}\\ So far, our discussion has mainly focused on the memory level. 
In future work, we will extend our analysis to fault-tolerant universal quantum computing with low overhead. 
We will investigate the implementation of fault-tolerant gates encoded in our tailored codes. 
A circuit level estimation of the error rates and a full analysis of the resource cost for fault-tolerant quantum computing will also be carried out. 

In Ref.~\cite{dua2022clifford}, it is shown that random Clifford deformations of the CSS surface codes can lead to better codes. 
It is worth investigating whether a similar transformation can boost the performance of our XZZX codes. 

It is also possible to generalize the construction of the XZZX GTCs to the recently advanced quantum low-density-parity-check (LDPC) codes \cite{hastings2021fiber, breuckmann2021balanced, panteleev2021quantum, panteleev2021asymptotically, breuckmann2021quantum, roffe2022bias}, which have asymptotically finite coding rate and good block-length-to-distance scaling. 
The current construction of the quantum LDPC codes focuses mainly on homological CSS codes. 
The results in our paper indicate that non-CSS and non-homological construction might lead to more efficient codes against both symmetric and asymmetric noise. 

\section{Methods}
\noindent
\textbf{Repetition structure of the GTCs}\\ In this section we provide a detailed analysis of the repetition structure of the GTCs. 
A GTC has a repetition-$Z$ ($X$) structure iff it is cyclic along (1,1) ((-1,1)) direction; A GTC has a repetition-$Y$ structure iff it is cyclic along (0,1) and (1,0) direction. 
In the Results, we claim that a GTC has a repetition-$Z$ ($X$) structure iff it is cyclic along (1,1) ((-1,1)) direction; A GTC has a repetition-$Y$ structure iff it is cyclic along (0,1) and (1,0) direction. 
We provide the proof in the following.

The proof for repetition-$Z$ ($X$) structure is straightforward. 
For infinite $Z$ ($X$) noise, the GTCs are effectively single or disjoint sets of repetition codes obtained by removing the Pauli $Zs$ ($Xs$) in the stabilizers. 
A GTC is a single repetition code iff there are no logical operators consisting of only Pauli $X$s ($Z$s) that have weight smaller than $n$. 
Since the Pauli $Z$ ($X$) chains lie along $(1,1)$ ($(-1,1)$) direction, the above condition is equivalent to that there are no sub-cycles along the $(1,1)$ ($(-1,1)$) direction. 
Next we prove that it is necessary for a GTC to be cyclic along $(1,0)$ and $(0,1)$ direction in order to have a repetition-$Y$ structure. 
Suppose the code is not cyclic along either $(1,0)$ or $(0,1)$ direction, i.e.\ there are sub-cycles along that direction, then a Pauli-$Y$ string associated with a sub-cycle is a logical operators (with weight smaller than $n$), contradicting the assumption of the  repetition-$Y$ structure. 
To prove the sufficiency, we first define a classical ``Y-code" with parity-check matrix $H$, where each row of $H$ is associated with a stabilizer generator and $H_{i,j}$ = 1 iff the action of $S_i$ on the $j$th qubit is non-identity. 
A GTC under pure $Y$ noise is then decoded as a Y code. 
Given the condition that a GTC is cyclic along both $(1,0)$ and $(0,1)$ direction, without loss of generality we can choose $\vec{L}_1, \vec{L}_2$ as $\vec{L}_1 = n(1,0), \vec{L}_2 = (-m,1)$, such that $\gcd(m, n) = 1$ (to ensure the code is cyclic along (0,1) direction). 
We then label the qubits along the $(1,0)$ direction and correspondingly, the $i$-th row of $H$ is: $H_{i,j} = 1$ for $j = i,i + 1, i+m, i+1+m (\mod n)$ and $0$ otherwise. 
This is equivalent to (up to regrouping the stabilizer generators) the repetition code with $H_{i,j} = 1$ for $j = i, i + 1$ since $m$ and $n$ are coprime.

~\\
\noindent
\textbf{Algebraic Description of non-two-colorable GTCs}\\
Formally, we can describe the non-two-colorable codes with a length-three chain complex:
\begin{equation}
   \mathcal{A}_{2} \stackrel{\partial_{2}}{\longrightarrow} \mathcal{A}_{1} \stackrel{\partial_{1}}{\longrightarrow} \mathcal{A}_{0},
\end{equation}
where $\mathcal{A}_{0}, \mathcal{A}_{1}, \mathcal{A}_{2}$ are $\mathbb{Z}_2$ vector spaces with dimension $n-1, 2n, n-1$ respectively, and  $\mathcal{A}_1 \simeq \mathcal{P}/Z(\mathcal{P}), \mathcal{A}_2 \simeq \mathcal{A}_0 \simeq S$. 
$\mathcal{P}$ denotes the Pauli group and $Z(\mathcal{P})$ denotes its center. $S$ denotes the stabilizer group.
We choose the basis of $\mathcal{A}_1$ as the symplectic representation of Pauli operators, i.e.\ $P_i \in \mathcal{P} \rightarrow \phi(P_i)$, where $\phi$ denotes the symplectic representation. 
And the $i$-th basis of $\mathcal{A}_2$ and $\mathcal{A}_0$ is chosen as the $i$-th stabilizer generator. 
Under such a basis choice, the boundary map is given by $\partial_1 = \phi(S) \Omega, \partial_2 = \phi(S)^T$, where $\phi(S)$ is the symplectic representation of stabilizer generators (or the check matrix) and $\Omega := \left[\begin{array}{cc}
0 & I_{n} \\
-I_{n} & 0
\end{array}\right]$. 
The boundary identity $\partial_1 \partial_2 = 0$ imposes the commutation relation between stabilizers $\phi(S) \Omega \phi(S)^T = 0$. 
Now the logical operators of the code are associated with the elements of the first homology group of $H_1 \equiv \mathrm{ker}(\partial_1)/\textrm{im}(\partial_2)$. 
Geometrically, we can associated the basis elements  of $\mathcal{A}_2, \mathcal{A}_1$ and $\mathcal{A}_0$ with the plaquetttes, edges and vertices of the doubled graph embedded on the doubled torus. 
In other words, a non-two-colorable code $\textrm{GTC}(\vec{L}_1, \vec{L}_2)$ is embedded on the doubled torus with the doubled periodicity basis vectors $\vec{L}_{1,d}, \vec{L}_{2,d}$. 
Therefore, $H_1 \simeq \mathbb{Z}_2 \times \mathbb{Z}_2$ and there are three logical operators $Z_L, X_L, Y_L$ associated with three homologically nontrivial loops on the doubled torus. 
We note that these non-two-colorable codes should be distinguished with from the conventional homological codes since their logical operators are only associated with the first homological group, instead of the direct sum of the first homology and cohomology group (which requires the well-defined homology-cohomology duality) of the standard chain complex~\cite{kitaev2003fault, bravyi1998quantum, bombin2007homological, breuckmann2021quantum}. 
See Supplementary Material~\cite{SM} for more details.

~\\
\noindent
\textbf{Calculation of the effective code distance/half distance}\\
In this section, we provide an efficient algorithm to calculate the effective distance (under an independent XZ model) $d^{\prime}$, as well as an improved estimation of the modified half distance $t^{\prime}$.

The following algorithm takes the bias parameter $\omega$, doubled periodicity vectors $\vec{L}_{1,d}, \vec{L}_{2,d}$, the accuracy parameter $\epsilon$ as input , and outputs the effective code distance $d^{\prime}$ within accuracy $\epsilon$. 
Here $\round{\cdot}$ is the function that rounds to the nearest integer. 
The complexity of this algorithm is $O(d^{\prime 2}/\epsilon)$.

\begin{algorithm}[ht!]
    \caption{Algorithm for calculating $d^{\prime}$}\label{euclid}
    \hspace*{\algorithmicindent} \textbf{Input:} $\omega, \vec{L}_{1,d}, \vec{L}_{2,d}$, $\epsilon$\\
    \hspace*{\algorithmicindent} \textbf{Output:} $d^{\prime}$ 
    \begin{algorithmic}[1]
    \State 
 $\vec{L}_{i,d}^{\prime} \gets \vec{L}_{i,d}\begin{bmatrix}
   \omega & 0\\ 
   0 & 1 \\
   \end{bmatrix}$ for $i = 1,2$, $L \gets (\vec{L}_{1,d}^{\prime T},\vec{L}_{2,d}^{\prime T})^T$ \Comment{$\vec{L}_{i,d}$ are row vectors, $L$ is a $2\times 2$ matrix}
    \State $r \gets 0, s \gets 0$  
    \While{$s = 0$}
        \State $r \gets r + \frac{\epsilon}{2}$
        \For{each $k \in [0, \frac{2r}{\epsilon}]$}
            \State $x,y \gets -r + k\frac{\epsilon}{2}, k\frac{\epsilon}{2}$
            \If{$\|L\round{L^{-1}(x, y)^T} - (x, y)^T\|_1 \leq \epsilon$} \Comment{$\round{\cdot}$ rounds the entries of a vector to integers.}
                \State $s \gets 1$
            \EndIf
            \State $x,y \gets r - k\frac{\epsilon}{2}, k\frac{\epsilon}{2}$
            \If{$\|L\round{L^{-1}(x, y)^T} - (x, y)^T\|_1 \leq \epsilon$}
                \State $s \gets 1$
            \EndIf
        \EndFor
    \EndWhile
    \State $d^{\prime} \gets r$
\end{algorithmic}
\end{algorithm}


To more accurately characterizes how the logical error rate scales with the (total) physical error rate, we define the following effective half code distance: For an asymmetric Pauli channel with probability distribution $\{p_{\sigma} \}, \sigma \in \{X,Y,Z\}$ and total error probability $p$, the effective half distance $r^{\prime}$ of a code is the minimum modified weight of any uncorrectable errors, with the noise-modified weight of a Pauli $\sigma$ given by: $\textrm{wt}^{\prime}(\sigma) \equiv \log p_{\sigma}/(\max_{\sigma}\log p_{\sigma})$.

With the above defined $r^{\prime}$, the logical error rate (to the leading order) scales as $p_L \propto p^{r^{\prime}}$. 
Note that for the depolarizing noise
$r^{\prime}$ is simply given by $r^{\prime} = \lfloor{(d^{\prime} + 1)/2}\rfloor$, which can be efficiently calculated by if $d^{\prime}$ is known. 
However, for an asymmetrical Pauli channel, $r^{\prime}$ not necessarily equals $\lfloor(d^{\prime} + 1)/2\rfloor$ and can not be efficiently calculated in general. 
Instead of approximating $r^{\prime}$ as $\lfloor{(d^{\prime} + 1)/2}\rfloor$ in Fig.~\ref{fig:verification_of_effective_code_distance}, which fails in some cases, we can adopt a better approximation of $r^{\prime}$:
(1) Find the logical operator $L_m$ with the minimum effective weight. 
(2) Approximate $r^{\prime}$ as $r^{\prime} \approx \min_{E \subset L_m} \textrm{wt}^{\prime}(E)$, where the minimization is over all the subsets of the logical operator. 
The above calculation can be done efficiently provided that $L_m$ can be efficiently located (or equivalently, $d^{\prime}$ can be calculated efficiently). 

~\\
\noindent
\textbf{Decoders}\\
In this section, we present the details of the decoders, including the MWPM decoder and the TN decoder, that we use in the main text. 
The decoding/recovery problem for an $[[n,k,d]]$ quantum stabilizer code $\mathcal{S}$ is as follows. 
Given a syndrome $\vec{s} \in \{0,1\}^{n-k}$ obtained from the stabilizer measurements, we identify the possible errors and correspondingly apply a correction $R \in \mathbb{P}_n$. 
If an error $E \in \mathbb{P}_n$ occurred, the recovery is successful only if $R E \in \mathcal{S}$.

Minimum-weight perfect matching (MWPM) finds the most likely error pattern given a syndrome $\vec{s}$ by matching the defects by pairs on a complete weighted graph. 
The complete graph is constructed by assigning syndrome defects to the vertices and choosing the weights of the edges according to the error probabilities of the possible errors that create the associated defect pairs. 
The efficient algorithm due to Edmond \cite{edmonds1965paths} returns a perfect matching of the graph such that the sum of the weights of the edges of the matching is minimal. 
The returned matching can then be used to apply the corrections. 

The success of the MWPM decoder depends on how the weights of the edges of the input graph are assigned, which we specify here. 
The assignment follows Ref.~\cite{o2017density}. 
First, we construct a weighted ancilla graph $\mathcal{G}_A = (V_A, E_A)$, in which each vertex is associated with a stabilizer generator and two vertices $u, v$ are connected by an edge $e_{uv}$ if and only if a \textrm{single} Pauli error $P_{e_{uv}}$ ($X$ or $Z$ for our XZZX GTCs) creates two defects on stabilizers associated with $u$ and $v$. 
The weight of an edge is assigned as the probability of the associated (single) Pauli error. 
Let $A_A$ be the weighted adjacency matrix on $\mathcal{G}_A$, i.e.\ $(A_A)_{uv} = p(P_{e_{uv}})$. 
We then obtain a full-connected syndrome graph $\mathcal{G}_S$, which has the same vertices as $\mathcal{G}_A$ but with full connectivity. 
To assign the weight for $\mathcal{G}_S$, we first calculate the following $A_S$ matrix:
\begin{equation}
    A_{S}=A_{A}+A_{A}^{2}+A_{A}^{3}+\cdots=\frac{1}{1-A_{A}}-1.
    \label{eq:adjacency_mat}
\end{equation}
Eq.~\eqref{eq:adjacency_mat} gives the $(u, v)$ entry of $A_S$:
\begin{equation}
\begin{aligned}
    (A_S)_{uv} & = \sum_{(e_1, e_2, ..., e_n) \in \mathcal{P}_{u,v}} \prod_{j=1}^{n} (A_A)_{uv} \\
    & = \sum_{(e_1, e_2, ..., e_n) \in \mathcal{P}_{u,v}} \prod_{j=1}^{n} \textrm{Pr}(P_{e_j}), \\
\end{aligned}
\end{equation}
where $\mathcal{P}_{u,v}$ denotes all the paths between $u$ and $v$. 
$(A_S)_{u,v}$ is approximately the sum over the probability of all possible error chains producing the defects $u,v$. 
We then assign the weight of the edge connecting $u,v$ in $\mathcal{G}_S$ as $\textrm{wt}(u,v) = - \log (A_S)_{uv}$. 
Then during the QEC, when a syndrome $\vec{s}$ with a set of defects $\mathcal{E}$ is measured, a subgraph of $\mathcal{G}_S$ containing only the defect vertices $\mathcal{E}$ is used as the input graph for the matching algorithm.

We note that the MWPM algorithm with the above weight assignment is close to optimal under the independent XZ model. 
However, it is sub-optimal under the biased noise model which assumes that the $Y$ and $X$ error happens with equal probability, due to its inability to handle the correlation between $X$ and $Z$ errors.

Next, we present the details for the tensor network (TN) decoder. 
Given a syndrome $\vec{s}$, there exists a whole class of Pauli operators consistent with the syndrome. 
If $f(\vec{s})$ is a syndrome consistent Pauli, the cosets $f(\vec{s}) \mathcal{S}$, $f(\vec{s}) X_L \mathcal{S}$, $f(\vec{s}) Z_L \mathcal{S}$, and $f(\vec{s}) Y_L \mathcal{S}$ enumerate all Pauli operators consistent with $\vec{s}$. 
Here $\mathcal{S}$ denotes the stabilizer group and $X_L, Y_L, Z_L$ denote three logical operators.
The decoding problem finds the most probable coset, and outputs any Pauli in that coset. 
Brute force computation of coset probabilities has an exponential overhead cost in the number of qubits, rendering such a method intractable for thousands of Monte Carlo iterations. 
However, such optimal decoding methods are desired to estimate the best case performance of quantum codes. 
So here we describe a more tractable method for a class of XZZX GTCs by tensor network contraction, the BSV decoder \cite{bravyi2014efficient} adapted to GTCs,

For toric codes of the XZZX type, one can take advantage of the fact that the fundamental parallelogram has a certain freedom. 
For an GTC cyclic along $\hat{l}_0 = (1,0)$ or $\hat{l}_0 = (0,1)$, qubits can be uniquely labeled along a horizontal or vertical line. 
This means that the tensor network describing coset probabilities $\mathrm{prob}(E \mathcal{S})$ for Pauli error $E$ is a linear chain, with non-local coupling. 
Each tensor has rank 6 \cite{tuckett2019tailoring} where coupling between qubits along the direction $\hat{l}_0$ are of bond dimension 4, and non-local coupling is bond dimension 2. 
If $n \hat{l}_0$ is a lattice vector, then $\hat{l}_0^{\perp} + z \hat{l}_0$ is a lattice vector where $\hat{l}_0^{\perp} \cdot \hat{l}_0 = 0$ and $z \in \mathbb{Z}$. 
For example, the $[[13,1,5]]$ code with $\vec{L}_1 = (3,2)$, $\vec{L}_2 = (-2,3)$, $\vec{L}_1 + \vec{L}_2 = (1,5)$ so $|z| = 5$. 
$|z|$ characterizes the non-locality, so higher $|z|$ corresponds to a more costly scheme. 
The contraction scheme prioritizes maximizing trace legs, first reducing network to a chain without any non-local couplings, then contracting the rest. 
The very crude upper bound on complexity is the number of contractions times the complexity of the worst possible contraction step which ends up as $O(n \cdot 4^{|z|})$. 
For codes with higher $|z|$, the corresponding tensor network decoding scheme is harder to contract. 
The most non-local codes are axis-aligned, square toric codes (with $\vec{L}_1 = (a, 0), \vec{L}_2 = (0, a)$ for $a \in \mathbb{Z}$), where the corresponding tensor network is a trace of a projected entangled pair state (PEPS) form, which are hard to contract in general and unsuited for Monte Carlo techniques with many iterations. 
In contrast, the TNs for the optimal or close-to-optimal codes presented in the Results are relatively easy to contract. 

~\\
\noindent
\textbf{Fault tolerant quantum error correction using flag qubits}

In this section, we prove that we can use one flag qubit to fault-tolerantly operate the GTCs and realize fault-tolerant quantum error correction (FTQEC). 
Before the proof, we define some notations to facilitate the analysis.

Let $C(P)$ be a circuit that implements a projective measurement of a Pauli $P$ and does not flag if there are no faults.
Let $\mathcal A$ and $\mathcal B$ be two sets of Pauli operators. 
We define a new set of Pauli operators $\mathcal A \times \mathcal B$ as follows
\begin{equation}
\mathcal A \times \mathcal B = \{ AB | \forall A\in \mathcal{A}, B\in \mathcal{B}\}.   
\end{equation}
When we consider a circuit-level noise, we need to consider all the potential physical faults, including gate failures, idling errors, state preparation, and measurement errors. 
Since now we try to estimate how different faults contribute to the logical error rate, we similarly assign effective weights to the various faults according to their probability to occur.

We follow some of the definitions in Ref.~\cite{chamberland2018flag} and adapt them to our biased-noise case.  

\begin{definition}[$t^{\prime}$-flag circuit]
A circuit $C(P)$ is a $t^{\prime}$-flag circuit if the following holds: For any set of faults with effective weight $v^{\prime} \leq t^{\prime}$ in $C(P)$ resulting in an error $E$ with $\min(\textrm{wt}^{\prime}(E), \text{wt}^{\prime}(EP)) > v$, the circuit flags.
\end{definition}

\begin{definition}[flag error set]
Let $\mathcal{E}_{m^{\prime}}\left(g_{i_{1}}, \cdots, g_{i_{k}}\right)$ be the set of all possible data errors caused
by physical faults with total effective weight $m^{\prime}$ spread amongst the circuits $C\left(g_{i_{1}}\right), C\left(g_{i_{2}}\right), \cdots, C\left(g_{i_{k}}\right)$, which all flagged. 
\end{definition}

\begin{definition}[flag $t^{\prime}$-FTEC condition]
Let $\mathcal{S}=\left\langle g_{1}, \ldots, g_{r}\right\rangle$ be a stabilizer code and $\left\{C\left(g_{1}\right), \ldots, C\left(g_{r}\right)\right\}$ be a set of $t^{\prime}$ flag circuits . 
For any set of m stabilizer generators $\left\{g_{i_{1}}, \cdots, g_{i_{m}}\right\}$ such that $1 \leq m \leq t^{\prime}$, any pair of errors $E, E^{\prime} \in \mathcal{E} \equiv \bigcup_{j^{\prime}=0}^{t^{\prime}-m} \mathcal{E}_{t^{\prime}-j^{\prime}}\left(g_{i_{1}}, \cdots, g_{i_{m}}\right) \times \mathcal{E}_{j^{\prime}}$ satisfies $E E^{\prime} \notin \mathcal{C}(\mathcal{S})\backslash \mathcal{S}$, where $\mathcal{C}(\mathcal{S})$ denotes the centralizer of $\mathcal{S}$.
\label{condition:flag_FT}
\end{definition}
Here $\mathcal{E}_{j^{\prime}}$ denotes the set of arbitrary Pauli errors on the data qubits with effective weight $j^{\prime}$. 
There are the following two key ingredients for a flag-QEC scheme to be fault-tolerant.
(1) Detectability of bad errors that propagate from ancilla qubits to data qubits. 
This is achieved by using $t^{\prime}$-flag circuit where $t^{\prime} = (d^{\prime} - 1)/2$ for a code with effective distance $d^{\prime}$. 
(2) Distinguishability (up to stabilizers) between errors that are associated with the same flag pattern. 
This condition in general depends on both the code and the flag circuit, which is case specific and has to be checked given a code and flag circuits.

\begin{figure}[ht!]
    \centering
    \includegraphics[width = 0.5 \textwidth]{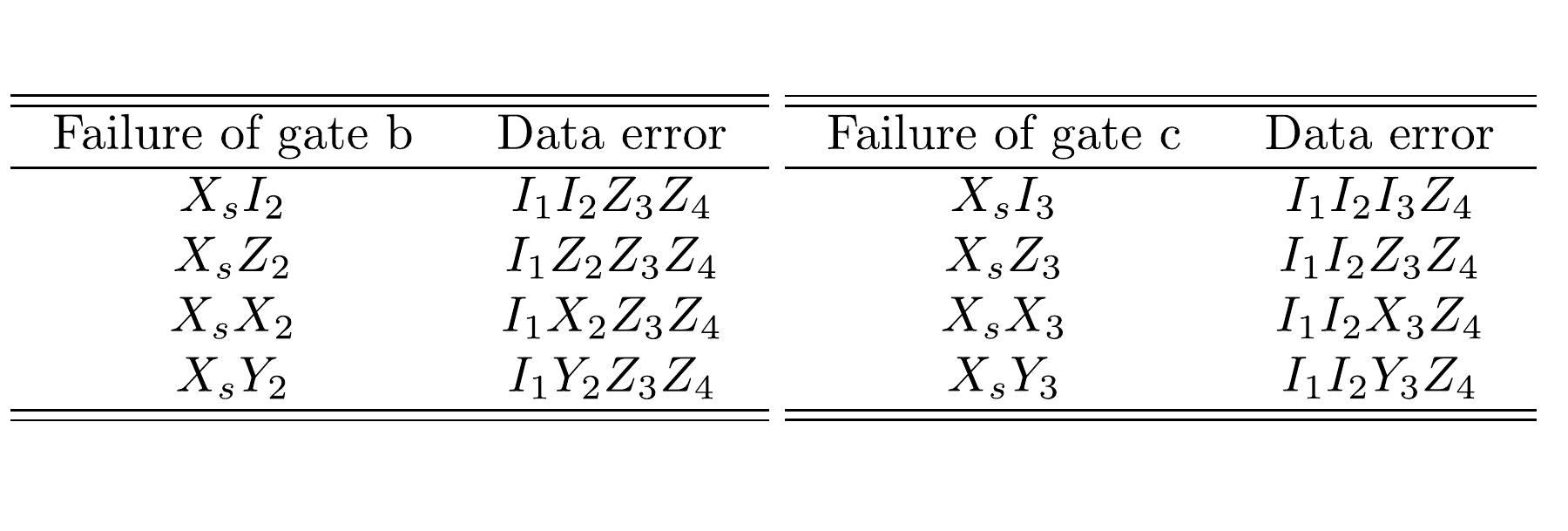}
     \caption{The propagation of bad gate errors that are caught by the flag qubit in Fig.~\ref{fig:flag_circuit} during the measurement of a stabilizer $X_1 X_2 Z_3 Z_4$. 
     We use the convention that the errors happen after an ideal gate. 
     Then only the failure of gates $b$ and $c$ can propagate into errors with larger effective weights on the data qubits (modulo the stabilizer to be measured). 
     For each table, in the left column, we list the possible failures of the gate $b$ (or $c$) that can trigger the flag, where $AB$ indicates an error $A$ on the control (syndrome qubit) and an error $B$ on the target (data qubits). 
     While in the right column we list the induced errors on the four data qubits (from top to down in Fig.~\ref{fig:flag_circuit}) on which the measured stabilizer is supported. 
     The set of listed gate errors are denoted by $\boldsymbol{X_s}$ and the resulted data errors are denoted by $\boldsymbol{\xi}$.}
    \label{fig:flag_error}
\end{figure}

To prove that the flag $t^{\prime}$-FTQEC condition is satisfied for the GTCs, we first specify the flag error set under consideration. 
For a flagged syndrome extraction circuit in Fig.~\ref{fig:flag_circuit}, we can classify possible physical faults by their components on the syndrome and flag qubits and consider the following set of single faults $\boldsymbol{E} = \boldsymbol{X_s} \cup \boldsymbol{Z_s} \cup \boldsymbol{X_f} \cup \boldsymbol{P}$ that are potentially harmful.
(1) $\boldsymbol{X_s}$: the set of faults that has a $X$ component on the syndrome qubit and can propagate to data errors with hamming weight larger than $1$. 
This set $\boldsymbol{X}_s$ can be viewed as (a subset of) errors of the gates b, c, listed in Fig.~\ref{fig:flag_error}. 
(2) $\boldsymbol{Z_s}$: the set of faults that has a $Z$ component on the syndrome qubit, which give wrong syndrome measurement outcome but neither propagate to data qubits nor trigger the flag. 
Therefore, $\boldsymbol{Z_s}$ can be suppressed by repeated syndrome extraction. 
This set includes $Z$ type of errors on the syndrome or flag qubits and measurement errors on the syndrome qubit. 
(3) $\boldsymbol{X_f}$: The set of faults with $X$ component on the flag qubit, which are not propagated from the syndrome qubit. 
This set include the $X$ errors or measurement errors on the flag qubit. 
$\boldsymbol{X_f}$ trigger the flags but do not propagate to data errors. 
(4) $\boldsymbol{P}$: a set of single Pauli errors on the data qubits within the support of the stabilizer to be measured. 
We note that a Pauli $Y$ error is considered to have both $Z$ and $X$ component. 

Since gate failures will induce errors on the data qubits, we need to use gates that do not convert low-effective-weight gate failures to data errors with higher effective weights in order to reach fault tolerance. 
We first assume that the gate errors listed in Fig.~\ref{fig:flag_error} all occur with a probability that equals the $p_X$ on the data qubits and they are all assigned with an effective weight $\omega$. 
This assumption is justified when we consider for example, the bias-preserving gates on stabilized cat qubits~\cite{chamberland2020building}. 
Under this noise model for the gates, the FTQEC condition will be satisfied when we consider the correlated $X$ and $Z$ noise model for the data qubits introduced in the Discussion, since the gate failures (with effective weights $\omega$) will introduce Pauli $Y$ errors on the data qubits. 
We note, however, if we assume that the gate failure $XY$ in Fig.~\ref{fig:flag_error} occurs with a smaller probability than $p_X$ and is assigned with effective weight $\omega + 1$, the FTQEC condition will also be satisfied under the independent $X$ and $Z$ noise model. 
In other words, to fault-tolerantly operate a code under a given noise model, we need to use gates that satisfy certain requirements.

We denote $\boldsymbol{p}(g_{i_j})$ as the set of possible faults triggering the flag during the measurement of the stabilizer $g_{i_j}$ (using flagged circuit). 
Furthermore, we denote $\boldsymbol{p}_1(g_{i_j}) \subset \boldsymbol{E}$ as the set of single faults triggering the flag and $\boldsymbol{p}_2(g_{i_j}) \subset \boldsymbol{E} \times \boldsymbol{E}$ as the set of double faults triggering the flag. 
And we denote $\boldsymbol{q}(g_{i_j}), \boldsymbol{q_1}(g_{i_j}), \boldsymbol{q_2}(g_{i_j})$ as the set of data errors caused by the corresponding faults. 
$\boldsymbol{p_i}(g_{i_j}), \boldsymbol{q_i}(g_{i_j})$ are summarized in Tab.~\ref{tab:pq_errors} for $i = 1,2$. 
We use $\boldsymbol{\xi} = \{IIZZ, IZZZ, IXZZ, IYZZ, IIIZ, IIXZ, IIYZ \}$ to denote the data errors resulted from $\boldsymbol{X}_s$ (shown in Fig.~\ref{fig:flag_error}). 
Here we omit the index of the data qubits for simplicity. 

\begin{table}[h]
    \centering
\begin{tabular}{|c|c|c|c|}
\hline
\multicolumn{2}{|c|}{ 1 fault } & \multicolumn{2}{c|}{ 2 faults } \\
\hline $\boldsymbol{p}_1$ & $\boldsymbol{q}_1$ & $\boldsymbol{p}_2$ & $\boldsymbol{q}_2$ \\
\hline  $\boldsymbol{X}_f$ & $I$ & $\boldsymbol{X}_f \times \boldsymbol{Z}_s$ & $I$ \\
\hline $\boldsymbol{X}_s$ & $\boldsymbol{\xi}$ & $\boldsymbol{X}_s \times \boldsymbol{Z}_s$ & $\boldsymbol{\xi}$ \\
\hline  & & $\boldsymbol{X}_f \times \boldsymbol{P}$ & $\boldsymbol{P}$ \\
\hline & & $\boldsymbol{X}_s \times \boldsymbol{P}$ & $\boldsymbol{\xi} \times \boldsymbol{P}$ \\
\hline
\end{tabular}
\caption{The set of data errors $\textbf{q}_1$ ($\textbf{q}_2$) induced by single physical faults $\textbf{p}_1$ (double faults $\textbf{p}_2$) during the measurement of a XZZX stabilizer using the circuit in Fig.~\ref{fig:flag_circuit}, with the flag triggered.}
\label{tab:pq_errors}
\end{table}

To prove that a GTC using the flag circuit in Fig.~\ref{fig:flag_circuit} satisfies the condition~\ref{condition:flag_FT}, it is sufficient to show that any error pair $E_{\textrm{pair}} = E E^{\prime}, E, E^{\prime} \in \mathcal{E}$ has effective weight smaller than $d^{\prime}$, therefore can not be a logical operator. 
We prove that this is the case when $E, E^{\prime} \in \mathcal{E}_{t^{\prime}}\left(g_{i_{1}}, \cdots, g_{i_{m}}\right)$ and the proof for the case when $E$ or $E^{\prime}$ are in $\mathcal{E}_{t^{\prime} - j^{\prime}}\left(g_{i_{1}}, \cdots, g_{i_{m}}\right) \times \mathcal{E}_{j^{\prime}}$ for $j^{\prime} \neq 0$ follows. 
We denote $p_E(g_{i_k})$ as the physical fault occurring during the measurement of the stabilizer $g_{i_k}$ that eventually contributes to $E$, and $q_E(g_{i_k})$ as the data error (a component of $E$) induced by $p_E(g_{i_k})$. 
By definition, $\sum_{k = 1}^{m} \textrm{wt}^{\prime}(p_E(g_{i_k})) = \sum_{k = 1}^{m} \textrm{wt}^{\prime}(p_{E^{\prime}}(g_{i_k})) = t^{\prime}$. 
If we can show that the $q_{E_{\textrm{pair}}}(g_{i_k}) \equiv q_{E}(g_{i_k}) q_{E^{\prime}}(g_{i_k})$ has effective weight no larger than $\textrm{wt}^{\prime}(p_{E}(g_{i_k})) +\textrm{wt}^{\prime}(p_{E^{\prime}}(g_{i_k}))$, then $\textrm{wt}^{\prime}(E_{\textrm{pair}}) \leq \sum_{k = 1}^{m} \textrm{wt}^{\prime}(q_{E_{\textrm{pair}}}(g_{i_k})) \leq 2t^{\prime} < d^{\prime}$ and consequently $E_{\textrm{pair}}$ can not be a logical operator. 
We show that this is true according to Tab.~\ref{tab:pq_errors} for the following two cases. 
(i) If $p_E, p_{E^{\prime}} \in \boldsymbol{p}_1$, we have $\textrm{wt}^{\prime}(p_E) + \textrm{wt}^{\prime}(p_{E^{\prime}}) = 2 \omega$. 
$q_{E_{\textrm{pair}}} \in \boldsymbol{\xi} \cup \boldsymbol{\xi} \times \boldsymbol{\xi}$. 
Note that \textit{elements in both $\boldsymbol{\xi}$ and $\boldsymbol{\xi}\times\boldsymbol{\xi}$ only have support on up to two qubits (modulo the stabilizers)}. 
As a result, $\textrm{wt}^{\prime}(q_{E_{\textrm{pair}}}) \leq 2 \omega$. 
(ii) If $p_E \in \boldsymbol{p}_1$ while $p_{E^{\prime}} \in \boldsymbol{p}_2$, we can similarly show that $\textrm{wt}^{\prime}(q_{E_{\textrm{pair}}}) \leq \textrm{wt}^{\prime}(p_E) + \textrm{wt}^{\prime}(p_{E^{\prime}})$. 
As an example, take $p_E \in \boldsymbol{X}_s$ and $p_{E^{\prime}} \in \boldsymbol{\xi}\times \boldsymbol{P}$, we have $\textrm{wt}^{\prime}(p_E) + \textrm{wt}^{\prime}(p_{E^{\prime}}) = 2 \omega + w_p$, where $w_p$ is the effective weight of that arbitrary Pauli error in $p_{E^{\prime}}$. 
Since $q_{E_{\textrm{pair}}} \in (\boldsymbol{\xi}\times \boldsymbol{\xi})\times \boldsymbol{P}$, we have $\textrm{wt}^{\prime}(q_{E_{\textrm{pair}}}) \leq 2 \omega + w_p = \textrm{wt}^{\prime}(p_E) + \textrm{wt}^{\prime}(p_{E^{\prime}})$. 
Note that we can easily see that $\textrm{wt}^{\prime}(q_{E_{\textrm{pair}}}) \leq \textrm{wt}^{\prime}(p_E) + \textrm{wt}^{\prime}(p_{E^{\prime}})$ also satisfies in other cases, where $p_E$ or (and) $p_{E^{\prime}}$ involves more faults. 
Till here we finish the proof. 
We note that similar proof applies for the independent $X$ and $Z$ noise model, if use gates whose $XY$ failure listed in Fig.~\ref{fig:flag_circuit} is of effective weight $\omega + 1$. 
In other words, our proposed FTQEC scheme for the GTCs using one flag qubit works under both independent and correlated Pauli $X$ and $Z$ noise model, assuming that appropriate gates with required failure rates are used.

\begin{acknowledgments}
We thank Senrui Chen, Arpit Dua, Michael Gullans, Ming Yuan, Pei Zeng  for helpful discussions. 
We are grateful for the support from the University of Chicago Research Computing Center for assistance with the numerical simulations carried out in this paper.
We acknowledge support from the ARO (W911NF-18-1-0020, W911NF-18-1-0212), ARO MURI (W911NF-16-1-0349, W911NF-21-1-0325), AFOSR MURI (FA9550-19-1-0399, FA9550-21-1-0209), AFRL (FA8649-21-P-0781), DoE Q-NEXT, NSF (OMA-1936118, EEC-1941583, OMA-2137642), NTT Research, and the Packard Foundation (2020-71479). 
A.S. is supported by a Chicago Prize Postdoctoral Fellowship in Theoretical Quantum Science.
\end{acknowledgments}

\bibliographystyle{apsrev4-1}

\end{document}


\title{Supplementary Material}

\author{Qian Xu}
\affiliation{Pritzker School of Molecular Engineering, The University of Chicago, Chicago 60637, USA}

\author{Nam Mannucci}
\affiliation{Pritzker School of Molecular Engineering, The University of Chicago, Chicago 60637, USA}

\author{Alireza Seif}
\affiliation{Pritzker School of Molecular Engineering, The University of Chicago, Chicago 60637, USA}

\author{Aleksander Kubica}
\affiliation{AWS Center for Quantum Computing, Pasadena, CA 91125, USA}
\affiliation{California Institute of Technology, Pasadena, California, 91125, USA}

\author{Steven T. Flammia }
\affiliation{AWS Center for Quantum Computing, Pasadena, CA 91125, USA}
\affiliation{California Institute of Technology, Pasadena, California, 91125, USA}

\author{Liang Jiang}
\email{liang.jiang@uchicago.edu}
\affiliation{Pritzker School of Molecular Engineering, The University of Chicago, Chicago 60637, USA}
\affiliation{AWS Center for Quantum Computing, Pasadena, CA 91125, USA}

\maketitle

\tableofcontents

\section{\label{sec:A_map} Map between XZZX cyclic codes and XZZX generalized toric codes (GTCs)}
In this section we explicitly construct the map from the XZZX cyclic codes to the XZZX GTCs. More specifically, given a XZZX cyclic code $\mathcal{S}(n, a, b)$, we try to find the corresponding $\textrm{GTC}(\vec{L}_1, \vec{L}_2)$ with $\vec{L}_1, \vec{L}_2$ determined by the parameters $n,a,b$.

For a GTC that can be mapped to a cyclic code, it has to be cyclic along a certain direction $\hat{l}_1$, along which qubits are labeled. 
Suppose that this is the case and the two periodicity vectors of the GTC are given by $\vec{L}_1 = n\hat{l}_1, \vec{L}_2 = \hat{l}_2$, where $\hat{l}_1, \hat{l}_2$ are non-parallel vectors with coprime coordinates that satisfy $|\hat{l}_1 \times \hat{l}_2| = 1$. Without loss of generality, we consider the case when a stabilizer $Z_{-a} X_0 X_b Z_{a + b}$ (all the indices are modulo $n$) of the $\mathcal{S}(n, a, b)$ is mapped to a ZXXZ stabilizer of the GTC supported on four qubits with coordinates $(0,1), (0,0), (1,1)$ and $(1,0)$, respectively. Since the qubits are labeled along the $\hat{l}_1$ direction, the $Z_{-a} X_0 X_b Z_{a + b}$ stabilizer should also be supported on qubits with coordinates $-a\hat{l}_1, (0,0), b\hat{l}_1,$ and $(a + b)\hat{l}_1$, respectively. Hence, the points $-a\hat{l}_1$ and $(0,1)$ should be identified (by the boundary conditions specified by $\vec{L}_1$ and $\vec{L}_2$), and the same goes for the points $b\hat{l}_1$ and $(1,1)$. Considering the above conditions, the map from $n, a, b$ to $\hat{l}_1, \hat{l}_2$ is then given by the following constrained equations:
\begin{equation}
\begin{array}{c}
-a \hat{l}_1 \sim (0,1) \\
b \hat{l}_1 \sim (1,1) \\
|\hat{l}_1 \times \hat{l}_2| = 1
\end{array}
\end{equation}
where $\vec{A} \sim \vec{B}$ means that $\vec{A}$ and $\vec{B}$ are identified on the torus defined by $\vec{L}_1, \vec{L}_2$, i.e.\ $\vec{A} - \vec{B} = m_1 \vec{L}_1 + m_2 \vec{L}_2$ for $m_1, m_2 \in \mathbb{Z}$. 
We conjecture that the solution to the above nonlinear Diophantine equations always exists. 
Instead of proving the conjecture in general, we construct the explicit solution in some special cases below.

(1): If $\gcd (n, b) \mid (a + 1)$ ($\gcd (n, b)$ devides $(a + 1)$), we can choose $\hat{l}_1 = (m, m + 1), \hat{l}_2 = b \hat{l}_1 - (1,1)$, where $m$ is the solution of $b m - n k = a + 1$ with variables $m, k$.

(2): If $\gcd (n, a) \mid (b + 1)$, we can choose $\hat{l}_1 = (1, m), \hat{l}_2 = - a \hat{l}_1 - (0,1)$, where $m$ is the solution of $a m - n k = b + 1$ with variables $m, k$.

An alternative formulation of the mapping is to instead focus on the labeling of qubits. Given a cyclic code $\mathcal{S}(n,a,b)$, we can label qubits associated with a plaquette stabilizer in the following way. 
The qubits at $(x,y), (x,y+1), (x + 1, y)$ and $(x + 1, y + 1)$ are labeled $i, i - a, i + a + b$ and $i + a$ ($\mod n$), respectively.
We note that the label increases by $ - a$ ($\mod n$) by moving one step vertically on the lattice, and the label increases by $a + b$ by moving one step horizontally. The solutions $\alpha,\beta \in \mathbb{Z}$ admitted by the modular equation
\begin{equation}
    \Big[ (a + b) \alpha - a \beta \Big] \mod n = 0
\end{equation}
form all possible periodicity vectors $\vec{L} = (\alpha,\beta) \in \textrm{span}(\vec{L}_1, \vec{L}_2)$ of the GTC corresponding to $\mathcal{S}(n,a,b)$.

As pointed out in the main text, the map from the XZZX cyclic codes to the XZZX GTCs is injective. We verify this by showing that only a subset of GTCs is cyclic. A GTC is cyclic if and only if there exists a direction along which the GTC is cyclic, which gives the following condition for a GTC being cyclic:
\begin{definition}
Cyclic condition for the GTCs: A $\textrm{GTC}(\vec{L}_1, \vec{L}_2)$ is cyclic iff $\exists \vec{L} = (L_x, L_y) \in \text{span}(\vec{L}_1, \vec{L}_2)$ s.t. $\gcd(L_x, L_y) = 1$.
\label{def:cyclic_condition}
\end{definition}
We first prove the sufficiency. We say that a vector $(x,y) \in \mathbb{Z}^2$ is a coprime vector if $x$ and $y$ are coprime. If a coprime periodicity vector $\vec{L}$ exists, we first find a direction $\hat{l}_0$ (which is a coprime vector) that satisfies $|\hat{l}_0 \times \vec{L}| = 1$ and then let $L_1^{\prime} = n \hat{l}_0, \vec{L}_2^{\prime} = \vec{L}$. Then the code $\textrm{GTC}(\vec{L}_1^{\prime}, \vec{L}_2^{\prime})$ defines the same GTC, which is cyclic along the $\hat{l}_0$ direction. We prove the necessity by showing a contradiction. Suppose that a $\textrm{GTC}(\vec{L}_1, \vec{L}_2)$ is cyclic, but does not have coprime periodicity vectors. The condition for the GTC being cyclic implies that there exists $\vec{L}_1^{\prime}, \vec{L}_2^{\prime} \in \textrm{span}(\vec{L}_1, \vec{L}_2)$ that satisfies $\vec{L}_1^{\prime} = n\hat{l}_0, |\vec{L}_2^{\prime}\times \hat{l}_0| = 1$ ($\hat{l}_0$ is a coprime vector). Then $\vec{L}_2^{\prime}$ has to be coprime since $|\vec{L}_2^{\prime}\times \hat{l}_0| = 1$, which contradicts the assumption.

Obviously, there are codes that do not satisfy the condition~\ref{def:cyclic_condition}, thus not cyclic. For example, the $\textrm{GTC}((0,p), (p, 0))$ does not have any coprime lattice vectors for $p \geq 2$ and as a result, these square-lattice toric codes are not cyclic.

\section{\label{sec:algebric_description} Algebric description of the GTCs}

In this section, we give a unified algebraic framework to describe both the two-colorable and the non-two-colorable GTCs, and more generally, both the homological and the non-homological quantum codes, and draw the distinction between them. 
A $[[n, k, d]]$ CSS quantum code constructed from two classical codes with parity check matrices $H_X$ and $H_Z$ can be described by a length-three chain complex~\cite{kitaev2003fault, bravyi1998quantum, bombin2007homological, breuckmann2021quantum}, which is a collection of three $\mathbb{Z}_2$ ($\{0, 1\}$) vector spaces $\{\mathcal{A}_i\}$ and some linear maps $\{\partial_i: \mathcal{A}_i \rightarrow \mathcal{A}_{i - 1}\}$ called ``boundary operators" between them. The vectors in $\mathcal{A}_2, \mathcal{A}_1$ and $\mathcal{A}_0$ are associated with the $Z$ stabilizers, the $n$-qubit Pauli-$X$ operators and the $X$ stabilizers, respectively. With a proper choice of basis, the boundary maps are given by the parity-check matrices: 
\begin{equation}
   \mathcal{A}_{2} \stackrel{\partial_{2} = H_Z^{\intercal}}{\longrightarrow} \mathcal{A}_{1} \stackrel{\partial_{1} = H_X}{\longrightarrow} \mathcal{A}_{0}.
   \label{eq:CSS_chain_complexes}
\end{equation} 
The boundary maps satisfy $\partial_1 \partial_2 = 0$ since $H_X H_Z^{\intercal} = 0$, which is guaranteed by the commutation relation between the $X$ and $Z$ stabilizers. The dimension of the vectors spaces satisfy $\textrm{dim}(\mathcal{A}_1) = n$ and $\textrm{dim}(\mathcal{A}_2) = \textrm{dim}(\mathcal{A}_0) = n - k$ (since there are $n - k$ stabilizer generators in total). The kernal of $\partial_1$, $\textrm{ker}(\partial_1)$, is associated with the Pauli-Z operators that commute with all the $X$ stabilizers. The image of $\partial_2$, $\textrm{im}(\partial_2) \cong \textrm{row}(H_Z)$ (the row space of $H_Z$), is associated with all the Pauli-$Z$ operators that belong to the stabilizer group. Therefore, the first homology group, $H_1 := \textrm{ker}(\partial_1)/ \textrm{im}(\partial_2)$, is associated with the logical $Z$ operators of the code. Similarly, the logical $X$ operators are associated with the first cohomology group $H_1^{\prime} := \textrm{ker}(\partial_1^{\prime})/ \textrm{im}(\partial_2^{\prime})$, where the co-boundary maps are given by $\partial_1^{\prime} = H_Z$ and $\partial_2^{\prime} = H_X^T$. In combination, all the logical operators are associated with the direct sum between the first and the second cohomology group (the homology groups can be also viewed as $\mathbb{Z}_2$ vector spaces), i.e. $\mathcal{L} \simeq H_1 \oplus H_1^{\prime}$.
We can also describe some non-CSS codes, such as the planar XZZX surface codes~\cite{ataides2021xzzx} and the two-colorable GTCs, using the afore-constructed chain complex, as long as they are obtained from some CSS codes by applying local Clifford transformations. The check matrix (the symplectic representation of the stabilizer generators) $\phi(S)$ of such codes is in the form $\phi(S) = \left[\begin{array}{cc}
H_Z & 0 \\ 0 & H_X
\end{array}\right] T$, where $T$ is a symplectic matrix of dimension $2n$. Consequently, such non-CSS codes can still be described by the chain complex Eq.~\eqref{eq:CSS_chain_complexes}, except that the basis elements in $\{\mathcal{A}_i \}$ are no longer associated with pure $Z-$ or $X$-type operators. Instead, the basis is transformed by the local Clifford transformation represented by $T$. We call the above codes homological codes since they are described by the standard homological algebra --- the length-three chain complex Eq.~\eqref{eq:CSS_chain_complexes} (up to local Clifford transformation).

We can generalize the above construction for homological codes and describe any stabilizer quantum codes using the following length-three chain complex:
\begin{equation}
   \mathcal{A}_{2} \stackrel{\partial_{2} = \phi(S)^{\intercal}}{\longrightarrow} \mathcal{A}_{1} \stackrel{\partial_{1} = \phi(S) \Omega}{\longrightarrow} \mathcal{A}_{0}
   \label{eq:chain_complex_nonhomological}
\end{equation}
where $\mathcal{A}_0, \mathcal{A}_1, \mathcal{A}_2$ are now $\mathbb{Z}_2$ vector spaces with dimension $n - k, 2n, n-k$, respectively. Let $\mathcal{P}$ denotes the $n$-qubit Pauli group and $Z(\mathcal{P})$ denotes its center. Then $\mathcal{A}_1 \cong \mathcal{P}/Z(\mathcal{P})$ is associated with all the $n$-qubit Paulis up to phases. $\mathcal{A}_2 \cong \mathcal{A}_0 \cong S$ is associated with all the stabilizers.  $\phi(S) \in \mathbb{Z}_2^{(n-k)\times 2n}$ is again the binary symplectic representation of the generators of the stabilizer group $S$ and $\Omega := \left[\begin{array}{cc}
0 & I_{n} \\
-I_{n} & 0
\end{array}\right]$.
With these choices of boundary operators, $\mathrm{im}(\partial_2) = \mathrm{row}(\phi(S)) \cong S$ and $\mathrm{ker}(\partial_1) = \{x \in \mathbb{Z}_2^{2n} | \phi(S) \Omega x = 0 \} \cong \mathcal{C}(S)$ ($\mathcal{C}(S)$ denotes the centralizer of $S$) so the first homology group $H_1 = \mathrm{ker}(\partial_1) / \mathrm{im}(\partial_2) \cong \mathcal{C}(S)/S$ corresponds to logical operators. For homological codes, we have $\partial_1 = \left[\begin{array}{cc}0 & H_Z \\ H_X & 0\end{array}\right]$ and $\partial_2 = \left[\begin{array}{cc}H_Z^T & 0 \\ 0 & H_X^T\end{array}\right]$ (up to a basis transformation corresponding to a local Clifford transformation). As a result, we can have a decomposition of $\mathcal{A}_1 = \mathcal{A}_1^{Z}\oplus \mathcal{A}_1^{X}$ (each of dimension $n$), such that $\textrm{ker}(\partial_1) = \textrm{ker}(H_X) \oplus \textrm{ker}(H_Z)$ and $\textrm{im}(\partial_2) = \textrm{row}(H_Z) \oplus \textrm{row}(H_X)$. And consequently, the first homology group $H_1 = \textrm{ker}(H_X)/\textrm{row}(H_Z) \oplus \textrm{ker}(H_Z)/\textrm{row}(H_X)$ can also be decomposed into two disjoint subsets associated with vectors in $\mathcal{A}_1^Z$ and $\mathcal{A}_1^X$, respectively. Intuitively, this is nothing but combining the $\mathcal{A}_2$ and $\mathcal{A}_0$ (associated with $Z$ and $X$ stabilizers respectively) in Eq.~\eqref{eq:CSS_chain_complexes} into the vector space $\mathcal{A}_0 \cong \mathcal{A}_2$ in Eq.~\eqref{eq:chain_complex_nonhomological}, and combining the $Z-$ and $X-$ type Paulis into the full Pauli group up to phases, which is isomorphic to $\mathcal{A}_1$ in Eq.~\eqref{eq:chain_complex_nonhomological}. Because of this structure, the logical operators associated with the first homology group has a bipartite structure. However, there are some stabilizer codes in general, such as the non-two-colorable GTCs, that are not equivalent to some CSS codes up to local Clifford transformations. Consequently, the logical operators of these codes associated with $H_1$ do not have the bipartite structure. In other words, one can not find a bipartition of $\mathcal{A}_1$ such that any logical operator is only associated with a vector in one of the disjoint subspaces of $\mathcal{A}_1$. We call these non-homological codes. For the non-two-colorable GTCs that are embedded on the torus, the vectors in $\mathcal{A}_1$ are associated with paths on the doubled graphs that are embedded on the doubled torus. In this case, the non-bipartite structure of $H_1$ indicates that we can not find a bipartition of the doubled torus such that the nontrivial cycles associated with the logical operators have a bipartite embedding.

\bibliographystyle{apsrev4-1}

%